\documentclass[traditabstract]{aa}
\usepackage{graphicx}
\usepackage{txfonts}
\usepackage{epsfig}
\usepackage{natbib}
\usepackage{color}

\begin{document}

\title{Evolution of galactic discs: multiple patterns, radial migration and disc outskirts}

\author{I.~Minchev\inst{1}, 
B.~Famaey\inst{2,3}, 
A.C.~Quillen\inst{4},
P.~Di~Matteo\inst{5},
F.~Combes\inst{6},  
M.~Vlaji{\'c}\inst{1}, 
P.~Erwin\inst{7,8}
J.~Bland-Hawthorn\inst{9}
}

\institute{Leibniz-Institut f\"{ur} Astrophysik Potsdam (AIP), An der Sternwarte 16, D-14482, Potsdam, Germany
\email{iminchev1@gmail.com}
\and
Universit\'e de Strasbourg, CNRS, Observatoire Astronomique, 11 rue de l'Universit\'e, 67000 Strasbourg, France
\and
AIfA, University of Bonn, Germany
\and
Department of Physics and Astronomy, University of Rochester, Rochester, NY 14627
\and
Observatoire de Paris-Meudon, GEPI, CNRS UMR 8111, 5 pl. Jules Janssen, Meudon, 92195, France
\and
Observatoire de Paris, LERMA, CNRS, 61 avenue de L'Observatoire, 75014 Paris, France
\and
Max-Planck-Institut f\"{u}r extraterrestrische Physik, Giessenbachstrasse, D-85748 Garching, Germany
\and
Universit\"{a}ts-Sternwarte M\"{u}nchen, Scheinerstrasse 1, D-81679 M\"{u}nchen, Germany
\and
Anglo-Australian Observatory, P.O. Box 296, Epping, NSW 2121, Australia}

\abstract{
We investigate the evolution of galactic discs in N-body Tree-SPH simulations. We find that discs, initially truncated at three scale-lengths, can triple their radial extent, solely driven by secular evolution. At the same time, the initial radial metallicity gradients are flattened and even reversed in the outer discs. Both Type I (single exponential) and Type II (down-turning) observed disc surface-brightness profiles can be explained by our findings. We show that profiles with breaks beyond the bar's outer Lindblad resonance, at present only explained as the effect of star-formation threshold, can occur even if no star formation is considered. We explain these results with the strong angular momentum outward transfer, resulting from torques and radial migration associated with multiple patterns, such as central bars and spiral waves of different multiplicity. We find that even for stars ending up on cold orbits, the changes in angular momentum exhibit complex structure as a function of radius, unlike the expected effect of transient spirals alone. We show that the bars in all of our simulations are the most effective drivers of radial migration through their corotation resonance, throughout the 3~Gyr of evolution studied. Focussing on one of our models, we find evidence for non-linear coupling among $m=1, 2, 3$ and 4 density waves, where $m$ is the pattern multiplicity. In this way the waves involved conspire to carry the energy and angular momentum extracted by the first mode from the inner parts of the disc much farther out than a single mode could. We suggest that the naturally occurring larger resonance widths at galactic radii beyond four scale-lengths may have profound consequences on the formation and location of breaks in disc density profiles, provided spirals are present at such large distances. We also consider the effect of gas inflow and show that when in-plane smooth gas accretion of $\sim 5$~$M_\odot$/yr is included, the outer discs become more unstable, leading to a strong increase in the stellar velocity dispersion. This, in turn, causes the formation of a Type III (up-turning) profile in the old stellar population. We propose that observations of Type III surface brightness profiles, combined with an up-turn in the stellar velocity dispersions beyond the disc break, could be a signature of ongoing gas-accretion. The results of this study suggest that disc outskirts comprised of stars migrated from the inner disc would have relatively large radial velocity dispersions ($>30$~km/s at 6~scale-lengths for Milky Way-size systems), and significant thickness when seen edge-on. 
}

\titlerunning{Evolution of galactic discs}
\authorrunning{I. Minchev et al.}

\maketitle

\section{Introduction}

It has been recognized for a long time now that the radial distribution of the light of stellar discs of most spiral galaxies is well approximated by an exponential profile \citep[e.g.][]{freeman70}. The classical works of \citet{vanderkruit79} and \citet{vanderkruit81} have shown that some spiral, mostly late-type edge-on galaxies, have truncated discs (i.e., the surface-brightness profile shows a very sharp exponential fall-off beyond the truncation radius, typically 2-4 exponential scale-lengths). A number of recent studies looking into galaxies at a range of inclinations have actually found that galactic discs do not end at the truncation, but rather exhibit a change in the exponential scale-length of the light distribution \citep[e.g.][]{pohlen02,erwin05,pohlen06,hunter06}. It is now acknowledged that the surface-brightness profiles of the mass-carrying old stellar population in the outer discs can be a simple continuation of the inner exponential profiles (Type I), steeper (Type II) or even shallower (Type III) than the extrapolations of the inner profiles to large radii (see \citealt{joss05,ellis07} and \citealt{erwin08}, hereafter EPB08, and references therein).

The diversity of disc surface-brightness profiles is observed for both early-type and late-type disc galaxies. The frequency of different shapes of surface-brightness profiles appears to correlate, however, with galaxy morphological type. Whereas discs for which the outer surface-brightness profiles are steeper than the extrapolation of the inner profiles (Type II) are more frequent in late-type galaxies, the fraction of galaxies with shallower outer profiles (Type III) rises toward earlier morphological types (\citealt{pohlen06}, EPB08). Galactic discs with single exponential profiles (Type I) appear to be much rarer in late-type galaxies (EPB08). For the classical truncated surface-brightness profiles the breaks occur at smaller normalized radii in late-type galaxies than early-type galaxies \citep{pohlen06}. 

By extending the EPB08 sample of 66 barred galaxies with 47 unbarred ones, \cite{gutierrez11} made the first clear statements about the trends of outer-disc profile types along the Hubble sequence and their global frequencies. They found that Type I profiles are most frequently found in early-type discs, decreasing from one-third of all S0--Sa discs to 10\% of the latest type spirals. Conversely, Type II profiles increase in frequency with Hubble type, from $\sim25\%$ of S0 galaxies to $\sim80\%$ of Sd--Sm spirals. The fractions of Type I, II, and III profiles for all disc galaxies (Hubble types S0--Sm) were found to be 21\%, 50\%, and 38\%, respectively.

Cosmic environment could also play an important role in shaping disc profiles. In a recent study, \cite{erwin12} showed that while S0 field galaxies present statistically almost equal shares of Types I, II, and III, the Type II discs are missing for a sample selected from the Virgo cluster. If this were caused by environmental effects, the authors proposed two solutions: either something in the cluster (or protocluster) environment transforms Type II profiles into Type I, or else something prevents a Type I to Type II transition which is common in the field. 

Resolving the stellar contents of galaxies gives access to much fainter surface-brightness
levels and allows for the estimation of stellar population ages and metallicities \citep{joss05}. This could provide an insight on the processes giving rise to the properties of the outskirts 
of galactic discs. In their study of the low mass galaxy NGC~4244, \citet{dejong07} have 
shown that the break in the stellar count profiles occurs at the same radius for young, intermediate age, and old stars, independently from the height above the galactic plane. The low-mass disc galaxy M33 has a truncated disc \citep{ferguson07}. Measurements of the radial change of stellar ages of the dominant stellar populations show that the inner negative 
stellar age gradient reverses at the radius where the break of the surface-brightness 
profile is measured \citep{barker07a,williams09}. The metallicity gradient beyond the 
truncation radius is consistent however with the extrapolation of the metallicity 
gradient of the inner disc \citep{barker07b}. In NGC~300, a galaxy with comparable 
properties to those of M33, a single exponential surface-brightness profile extends out 
to ten scale lengths of the galaxy \citep{joss05}. Similarly to M33, the inner negative 
metallicity gradient of NGC~300 is found to extend out to 6-7 scale lengths, but beyond 
which the metallicity gradient flattens (\citealt{vlajic09}, and see also \citealt{vlajic11} for similar results in NGC~7793).

A number of scenarios have been investigated to explain the observed diversity of surface-brightness profiles seen in galactic discs, which can be separated into two categories. The first category advocates that the light profiles result from stars formed in the observed distribution, as a results of either a limit in the gas distribution \citep{vanderkruit87}, or by imposing a threshold for star formation \citep{kennicutt89}. \citet{elmegreen06} have shown that a double-exponential profile could result from a multicomponent star formation prescription, where turbulent compression in the outskirts of the disc allows for cloud formation and star formation despite subcritical densities.

The second category of models speculate that stars in the outer discs were deposited in those regions by dynamical processes. \citet{sellwood02} have invoked the idea of radial migration from transient spiral arms, where a large number of transients can mix stars at the corotation radius (CR) radially with the largest migration coinciding with the growth of strong spirals. Stars beyond the break in the exponential profile can thus be scattered outward from the inner disc. Since old stars have longer timescales to populate the outer regions of galactic discs, this leads to a change in the radial mean stellar age profile at the break radius and a positive age gradient in outer disc \citep{roskar08b}. The latter theoretical predictions have found good agreement with observations (e.g., \citealt{yoachim10,yoachim12}). By analyzing high-resolution hydrodynamical disc galaxy simulations \citet{sanchez09} have argued, however, that the change in the age gradient at the break radius may, alternatively, be due to a different star formation rate between regions inside the break radius and those outside. The efficiency of radial migration caused by transient spiral density waves is suspected to be significantly lower in low-mass discs than in Milky Way (MW)-like galaxies, as strong transient spirals are not expected to grow in such systems \citep{gogarten10}. More recent comparison between the migration seen in MW-size and low-mass systems was presented by \cite{radburn12}, where observational evidence of radial migration was found for NGC~7793.

Minor mergers are also expected to play a role in populating the outer regions of galactic discs \citep{penarrubia06, quillen09, bird12, purcell11}. In the MW, small satellites on radial, in-plane orbits can cause mixing in the outer disc \citep{quillen09} and thus account for the fraction of low-metallicity stars present in the solar neighborhood \citep{haywood08}. Such perturbations also create transient spirals, giving rise to structure in the velocity \citep{minchev09} and the energy-angular momentum space \citep{gomez12a}. Recently, \cite{gomez12b} related features in the energy distribution of the SEGUE G-dwarf stellar sample to disc perturbations by the Sagittarius dwarf galaxy, also showing that this MW satellite may be responsible for vertical density waves near the Sun \cite{gomez12c}. Given that the effect of minor mergers can be seen near the Sun, perturbations on the outer discs must be even larger due to the lower disc density. \citet{younger07} have demonstrated that the observed excess surface-brightness at large radii relative to the inner exponential profile in some galaxies could be the result of minor mergers. This requires, however, a significant gas supply in the primary disc and the encounter to be prograde with moderate orbital angular momentum. 

  \begin{table*}
      \caption[]{Galaxy modeling parameters}
         \label{parameters}
         \centering
         \begin{tabular}{lccccc}
            \hline\hline
            & gS0 &gSa & gSb & gSa$_{acc}$ & gSb$_{acc}$ \\
            \hline
            $M_{B}\ [2.3\times 10^9 M_{\odot}]$ & 10 & 10 & 5 & 10 & 5 \\
            $M_{H}\ [2.3\times 10^9 M_{\odot}]$ & 50 & 50 & 75 & 50 & 75\\
            $M_{s}\  [2.3\times 10^9 M_{\odot}]$ & 40 & 40 & 20 & 40 & 20\\
            $M_{g}/M_{s}$ & --& 0.1 & 0.2 & 0.1 & 0.2\\
            $r_{B}\ [\mathrm{kpc}]$ & 2 & 2 & 1 & 2 & 1\\
            $r_{H}\ [\mathrm{kpc}]$ & 10 & 10 & 12 & 10 & 12\\
            $a_{s }\ [\mathrm{kpc}]$  & 4 & 4 & 5 & 4 & 5\\
            $b_{s }\ [\mathrm{kpc}]$  & 0.5 & 0.5 & 0.5 & 0.5 & 0.5\\
            $a_{g }\ [\mathrm{kpc}]$  & -- & 5 & 6 & 5 & 6\\
            $b_{g }\ [\mathrm{kpc}]$ & -- & 0.2 & 0.2 & 0.2 & 0.2\\
             \hline
            $N_g$  & --          & 80\,000   & 160\,000 & 80\,000 & 80\,000\\
	    $N_{s}$ & 320\,000 & 240\,000 & 160\,000 & 80\,000 & 80\,000\\
	    $N_{DM}$ & 160\,000 & 160\,000 & 160\,000 & 80\,000 & 80\,000\\
	    \hline
	    Acc. radius [kpc] & -- & -- & -- & 17 & 12 \\
	    Acc. radial width [kpc] & -- & -- & -- & 10 & 5 \\
	    Acc. rate [M$_{\odot}$/yr] & -- & -- & -- & 5 & 5 \\
            \hline
         \end{tabular}
   \end{table*}

The increase of the fraction of galactic discs with antitruncated profiles toward earlier morphological types (EPB08), and the tendency of bars to be larger in early-type disc galaxies \citep[e.g.][]{chapelon99,erwin05} could be used to argue that the presence of a bar might play an important role in shaping the properties of the outskirts of galactic discs. \citet{mf10} and \cite{minchev11a} have shown that spiral structure interacting with a central bar can be an extremely efficient mechanism for radial mixing in galactic discs. Although angular momentum changes are most prominent in the vicinity of the CR radius of each individual perturber(as shown by \citealt{sellwood02}), the non-linear coupling between the bar and spiral waves (e.g., \citealt{tagger87,sygnet88}) makes this mechanism effective over the entire galactic disc. Being non-linear, this new way of mixing can be significantly more efficient at increasing the angular momentum than transient spirals alone and works with both short- and long-lived spirals \citep[see][for a detailed discussion]{mf10}. Subsequent studies of radial diffusion coefficients have confirmed the validity of bar-spiral coupling \citep{shevchenko11,brunetti11}. Recently, \cite{comparetta12} showed that, even if patterns are long-lived, radial migration can result from short-lived density peaks arising from interference among density waves overlapping in radius. In the near future, the most promising technique to put observational constraints on the migration mechanism at work in the MW is ``chemical tagging" \citep{freeman02} with, e.g., the HERMES survey, which will allow us to directly measure the spread of evaporated open clusters across the Galaxy as a function of age \citep{joss10}.

The effect of bars on galactic disc density profiles have been investigated in several previous numerical studies. \cite{debattista06} related the occurrence of Type II profiles in barred N-body discs to the bar strength. They have found that if the bar is large enough, a break forms roughly at the outer Lindblad resonance (OLR), which was interpreted as the result of bar-spiral coupling in their simulations. \cite{foyle08} extended this study further by exploring a more extensive parameter space of the dark and baryonic matter in galaxies, with combinations of the structural parameters that control the various observed profile types and how those profiles evolve over time. Both of these studies found that angular momentum redistribution leads to higher central mass density concentration, to the development of a two-component profile and the evolution of the inner disc scale length over time toward higher values. In both the above works the initial conditions comprised of very extended radial disc distributions (such as Type I profiles) in which the formation of a break in disc profiles was studied. Thus, they could not address the question of what happens when the disc is initially limited in extent, as it is most likely the case at the time the bar forms. 

In this paper we extend these very interesting results by examining the evolution of {\it initially truncated} disc profiles as may be the case in a more realistic, inside-out disc formation scenario. Detailed analyses of the disc dynamics resulting in angular momentum exchange are presented. We also study the effect on disc extension and break formation when initial gas component is present in the disc, as well as continuous gas inflow.

\section{Description of the simulations}
\label{sec:sims}

We study three main runs of isolated disc galaxies from the GalMer database \citep{dimatteo07}: the giant S0, Sa, and Sbc runs (hereafter gS0, gSa, and gSb). In GalMer, for each galaxy type, the initial halo and the optional bulge are modeled as Plummer spheres, with characteristic masses $M_H$ and $M_B$, and characteristic radii $r_H$ and $r_B$, respectively. Their densities are given by
\begin{equation}\label{halo}
\rho_{H,B}(r)=\left(\frac{3M_{H,B}}{4\pi r^3_{H,B}}\right)\left(1+\frac{r^2}{r^2_{H,B}}\right)^{-5/2}.
\end{equation}

On the other hand, the initial gaseous and stellar discs follow Miyamoto-Nagai density profiles with masses $M_{g}$ and  $M_s$, and vertical and radial scale-lengths given by $b_{g}$ and $a_{g}$, and $b_s$ and $a_s$, respectively:
\begin{equation}
\begin{array}{l}
\rho_{g,s}(R,z)=\left(\frac{b^2_{g,s} M_{g,s}}{4 \pi}\right) \times 
\frac{a_{g,s} R^2+\left(a_{g,s}+3\sqrt{z^2+b^2_{g,s}}\right)\left(a_{g,s}+\sqrt{z^2+b^2_{g,s}}\right)^2 }
{ \left[a_{g,s}^2+\left(a_{g,s}+\sqrt{z^2+b^2_{g,s}}\right)^2\right]^{5/2}\left(z^2+b^2_{g,s} \right)^{3/2} }.
\end{array}
\end{equation}

\begin{figure*}
\includegraphics[width=18cm]{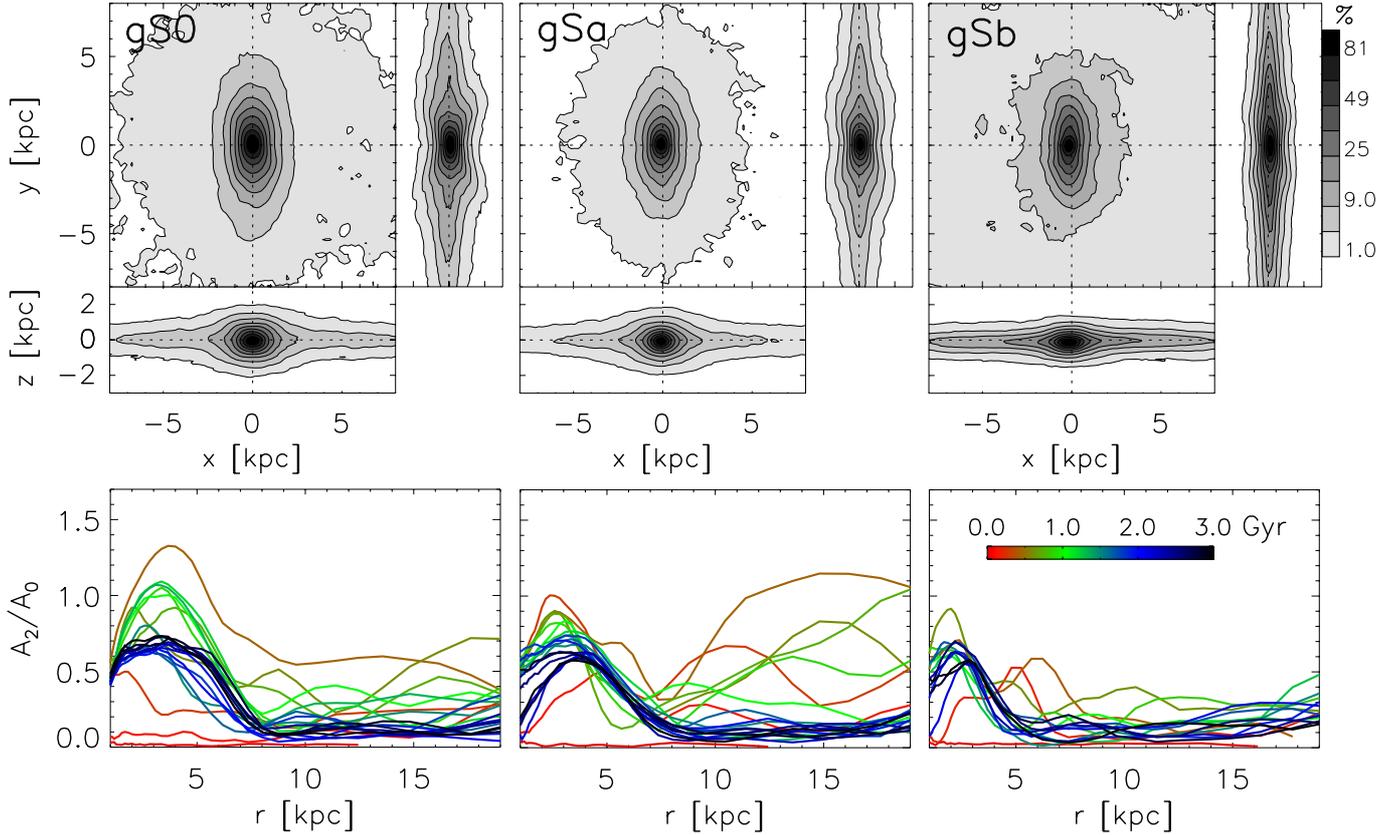}
\caption{
Top row: bar morphologies for the gS0, gSa, and gSb models at $t=2$~Gyr. For each galaxy we show a set of three panels: the disc surface density face-on view is presented in the square panel, with the bar side-on and head-on views in the attached right and bottom panels, respectively. Bottom row: $m=2$ Fourier amplitudes, $A_2/A_0$, as a function of radius from $t=0$ to $t=3$~Gyr as indicated by the color bar in the rightmost panel. The formation and evolution of a bar can be seen clearly in all inner discs. Deviations from zero in the outer disc indicate the strength of two-armed spirals.
\label{fig:bars}
}
\end{figure*}

The parameters for the initial conditions of the three runs, including the number of particles used for the gaseous disc, $N_g$, the stellar disc and bulge, $N_s$, and the dark matter halo, $N_{DM}$, are given in Table~\ref{parameters}. The initial Toomre parameter of both stars and gas is taken to be $Q=1.2$ as the initial condition of the Tree-SPH simulations and particle velocities are initialized with the method described in \cite{hern93}. Full details of the simulations are given in \cite{dimatteo07} and \cite{chilingarian10}.

Because the gaseous discs are set initially as Miyamoto-Nagai density distributions with arbitrary scale-heights, they are not in hydrostatic equilibrium. However, equilibrium is reached after the first 100-200~Myr. We have checked that the time evolution of the Toomre parameter for gS0 (dissipationless) and gSa (10\% gas) is similar, indicating that the instabilities developed in the gSa and gSb models described below are not the result of the initial gas instability.

\subsection{Gas and new stars}

Gas particles are ``hybrid particles" (SPH/new stars), characterized by two mass values: one corresponds to the total gravitational mass of the particle and remains unchanged during the entirety of the simulation, while the other is the gas content of the particle, decreasing or increasing according to the local star formation rate and mass loss (see \citealt{chilingarian10} for more details). Gravitational forces are always evaluated using the total gravitational mass, while hydro-dynamical quantities use the time-varying gas mass. When the gas fraction is below 5\% of the particle mass, it is transformed into a genuine star particle and the remaining gas mass is distributed over neighboring gas particles. We show the gas fraction as a function of radius at different times for the gSa and gSb models in Fig.~\ref{fig:gas_fr}. In this paper we do not discuss gas and newly formed stars separately, therefore,``hybrid particles" and ``gas/new stars" will be used interchangeably.

\subsection{Gas accretion}
\label{sec:acc}

In addition to the preassembled discs with an initial gas fraction described above, we also simulate the effect of external gas accretion. The infalling gas is deposited in an annulus in the disc plane, close to the edge of the gaseous disc. At this radius the gas is assumed to have already settled in the plane, moving with a circular velocity. To cope with the large mass variations of the gas component, the number of gas particles is kept constant, while the individual particle mass is varied. Unlike the GalMer simulation technique described above, where the gas particles are hybrid, here the mass which is subtracted from the gas for making stars is added to the neighboring disc stars. Therefore, in this prescription both the gas and stellar particles can increase in mass. For more details we refer the reader to \cite{combes11}. 

We simulate two galaxies with gas accretion: (1) gSa$_{acc}$, which starts with the same parameters as gSa, except for the gas being accreted at the constant rate of 5~$M_\odot$/yr in the annulus $r=17\pm5$~kpc and (2) gSb$_{acc}$, same as gSb initially, except for the gas being accreted at the constant rate of 5~$M_\odot$/yr in the range $r=12\pm2.5$~kpc. All parameters can be found in Table~\ref{parameters}.

\subsection{The bars in our simulations}

\begin{figure}
\includegraphics[width=8cm]{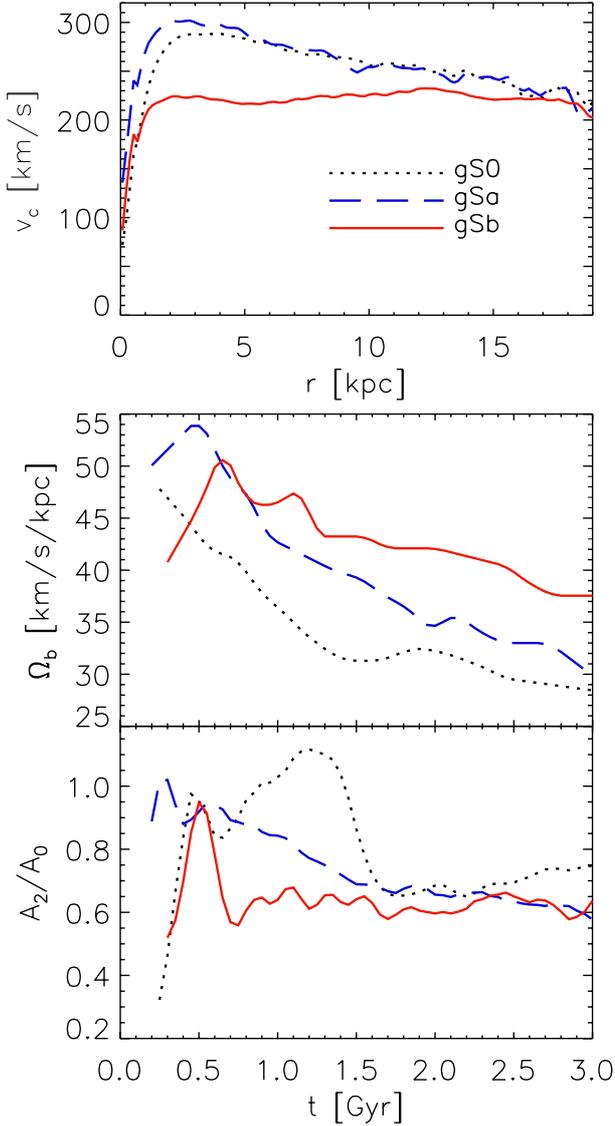}
\caption{
Top panel: circular velocities at $t=2$~Gyr for the gS0, gSa and gSb models. Middle panel: time evolution of the bar pattern speeds, $\Omega_b$. Bottom panel: $m=2$ Fourier amplitudes, $A_2/A_0$, as a function of time. Note that here we take the maximum values inside the bar; these are not sensitive to the lengths of the bars which can be seen in the bottom row of Fig.~\ref{fig:bars}.  
\label{fig:om}
}
\end{figure}

All models discussed in this paper develop central bars. Here we show the bars' general properties, such as morphology, strength and pattern speed, for the gS0, gSa and gSb models. Further analyses and relation to spiral structure are presented in Sec.~\ref{sec:power}.

The top row of Fig.~\ref{fig:bars} shows the bar morphology at $t=2$~Gyr. For each model we present a set of three panels: the disc surface density face-on view (square panel), the bar side-on (right) and head-on (bottom) views. To make these plots we only considered stars in the range shown in the figures, excluding the outer parts of the disc. The normalized contour separation levels we use are shown in the color bar in upper right panel. A peanut is formed in all three cases, quite weak for the weaker gSb bar. It is clear that the gS0 bar is the strongest of all, followed by the gSa. The gSb bar is weaker because the gas fraction is higher, and when the gas is driven to the center by the bar, it actually weakens the bar \citep{bournaud02,bournaud05}.

The bottom row of Fig.~\ref{fig:bars} shows the Fourier amplitude, $A_m/A_0$, of the density of stars (gS0) and stars + gas (gSa and gSb) as a function of radius, where $A_0$ is the axisymmetric component and $m$ is the multiplicity of the pattern; here we only show the $m=2$ component, $A_2/A_0$. Different colors indicate the time evolution of $A_2$ radial profile from $t=0$ to $t=3$~Gyr every 150~Myr. We estimate these by Fourier-analyzing the disc surface density distribution, as described in Sec.~4.11 by \cite{dimatteo07}. Deviations from zero indicate the presence of bi-symmetric structure, i.e., a central bar or 2-armed spirals. For example, for the gS0 model the bar is identifiable by the smooth curve in the inner disc, which at later times has a maximum at $\sim3$~kpc and drops almost to zero at $\sim8$~kpc. Deviations from zero seen beyond that radius are due to the spiral structure. The gS0 bar is the strongest for the first half of the simulation. At later times its maximum amplitude decrease and platoes in the range $r=(3,5)$~kpc, unlike the localized maximum at $r\sim3.5$~kpc in the case of gSa. The gSb bar is clearly much shorter than the other two, peaking at approximately the same $A_2/A_0$ value as gSa at later times. The time variation of the maximum Fourier amplitude values can be seen in Fig.~\ref{fig:om}, bottom panel. At the beginning of the simulation the spirals are strong, especially for gSa, decreasing in amplitude at later times (blue/black lines) as the disc heats up. At the same time both bars extend while their pattern speeds decreases significantly (see Fig.~\ref{fig:om}. This behavior of bar evolution is well known and is likely due to angular momentum loss to the inner halo \citep{athanassoula03}. Note, however, that some of the bar angular momentum is also transferred to the outer disc, resulting in the extended disc profiles seen in Fig.~\ref{fig:gSall}, discussed in the following section.

\begin{figure}
\centering
\includegraphics[width=9cm]{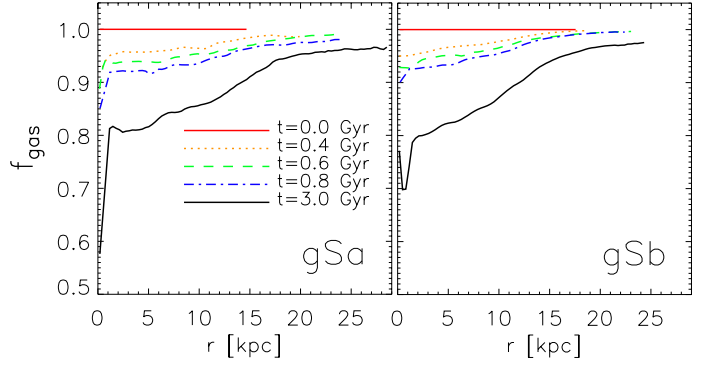}
\caption{
Gas fraction in the hybrid particles (initially 100\% gas) at different times for our dissipational simulations.
\label{fig:gas_fr}
}
\end{figure}

The top panel of Fig.~\ref{fig:om} presents the circular velocities for the gS0, gSa and gSb models in colors and line-styles as indicated in the figure. We estimate these from the disc stars only by correcting for asymmetric drift using the following equation \citep{bt08}:
\begin{equation}
V_{\rm c} = V^2_\phi + \sigma^2_\phi - \sigma^2_r \left(1+\frac{d({\rm ln}\,\nu)}{d({\rm ln}\, r) }\right) - r\frac{ d\sigma^2_r}{dr},
\end{equation}
where $V_{\rm c}$ is the circular speed and $V_\phi$ is the measured rotational velocity of stars for each bin. The initially identical circular velocities of gS0 and gSa differ at this time inwards of $\sim5$~kpc, with the gS0 starting to decline sooner as the galactic center is approached. This is likely due to the larger bar developed in the gS0 model. Both gS0 and gSa have declining circular velocities as a result of their large bulges. The gSb circular velocity is flat to large radii and similar to what is expected for the MW.

The middle panel of Fig.~\ref{fig:om} shows the time evolution of the bar pattern speeds, $\Omega_b$, for all models. For the dissipational gSa and gSb, $\Omega_b$ increases quickly after the bar forms and then slows down. This results from the mass increase inside the bar due to the gas inflow following the bar formation (see Fig.~\ref{fig:gSall}, third row). The gSb's shorter bar is reflected in its larger patterns speed. 

The bottom panel of Fig.~\ref{fig:om} shows the bar strengths as a function of time. We get these from the maxima of the $m=2$ Fourier amplitudes, $A_2/A_0$, inside the bar region. Further discussion of the bars' pattern speeds, strengths and relation to spiral structure can be found in Section \ref{sec:power}.

\begin{figure*}
\centering
\includegraphics[width=18cm]{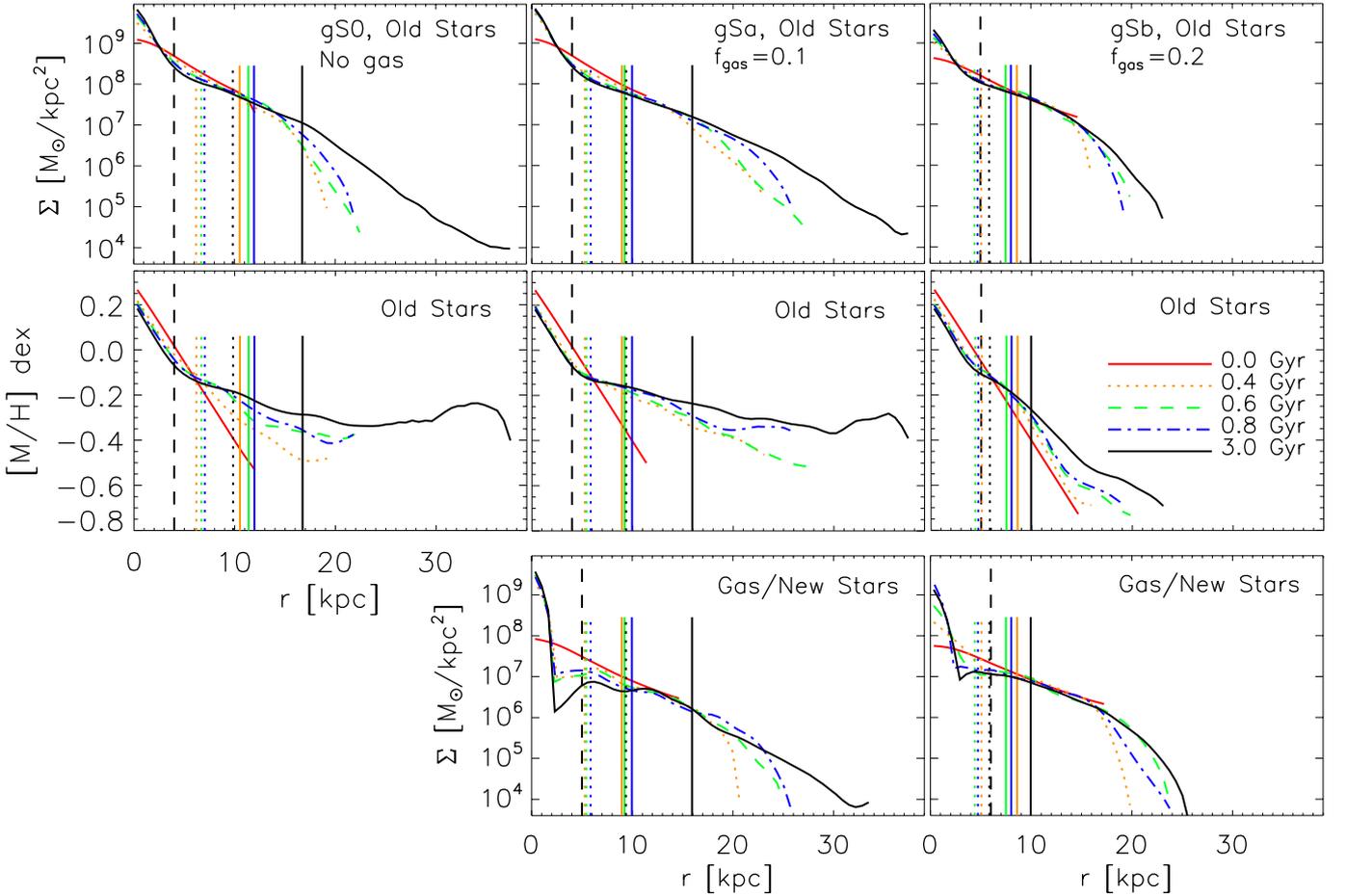}
\caption{
Temporal evolution of the azimuthally averaged density (top row) and metallicity (middle row) disc profiles for our galaxy models. Bulge particles are not considered. The different times shown are color coded as indicated in the second row, third column. The time evolution of the bars' corotation and outer Lindblad resonances are shown by the dotted and solid vertical lines, respectively, where the colors correspond to the different times. Note that the bars slow down considerably, especially in the simulation lacking a gaseous component (gS0).
\label{fig:gSall}
}
\end{figure*}

\section{Radial variation of the density and metallicity profiles}
\label{sec:gSall}

In Fig.~\ref{fig:gSall} we present the temporal evolution of the azimuthally averaged stellar disc surface density (first row) and metallicity (second row) radial profiles for our galaxy models. The third row presents the radial density of the gas and new stars. Bulge stars are not shown in order to compare the effect on the initially exponential discs for all galaxies. From top to bottom these are the gS0, gSa, and gSb models described in Sec.~\ref{sec:sims}. Densities are plotted logarithmically, thus an exponential appears as a straight line. Different lines in a given panel show each system at $t=0,40,60,80,$ and 3~Gyr. These are color-coded as indicated in the third row, rightmost panel. The vertical dashed, black lines show the initial scale-length, $h_d$, of each disc. The dotted and solid vertical lines give the radial positions of the bar's corotation and the OLR for each time indicated by the colors. The time evolution of the bar pattern speeds for all models is shown in Fig.~\ref{fig:om}. The initial gas fraction, $f_{\rm gas}$, is given at the top of each panel.

Examining the density profiles shown in the top row of Fig.~\ref{fig:gSall}, we observe that the initial disc particle distributions (red solid curves) expand well beyond the initial radius as the discs evolve. We note the appearance of two breaks in the initially single exponentials: (1) an inner one at $r\lesssim1.5h_d$ (near the bars' CR, dotted vertical lines), inwards of which the profiles steepen and (2) an outer one near the bars' OLR (solid vertical lines) beyond which the profiles also steepen. The metallicity profiles show similar changes with time, except that beyond the outer break the gradients reverse. There is, however, a number of remarkable differences among the evolution of the three galaxy model, which we now discuss.

\subsection{The effect of a gaseous component: the gS0 and gSa models}
\label{sec:gS0a}

Here we compare the gS0 and gSa simulations, which have the same initial conditions except for the lack of gas in the case of gS0. Inspecting the upper left panel of Fig.~\ref{fig:gSall} (gS0 model), we note that the disc, initially limited to 2.5$h_d$, expands to over 9$h_d$ in 3~Gyr. There is a break at $\sim13$~kpc for earlier times, which advances to $\sim17$~kpc at the final time, beyond which the exponential profile steepens. It is notable that this break does not occur at the initial truncation (12~kpc) and, thus, must be related to a dynamical process taking place in the disc, rather than the initial conditions. At times 0.6 and 3~Gyr, the profile outside the bar's CR can be fit by two exponentials reasonably well, with the break shifting from $\sim3h_d$ to $\sim4h_d$. This is consistent with the Type II (down-turning) profile, in the classification by EPB08, for example. Very interestingly, at $t=3$~Gyr (black solid curve) the profile between the bar's CR and the outer break appears as a single exponential $\sim1.5$ scale-lengths beyond the initial truncation radius. At this time, the break coincides exactly with the bar's OLR, therefore providing a perfect example of the Type II.o-OLR profile (see figure 5 by EPB08), where the break in the surface-brightness profile happens at an outer ring associated with the bar's OLR (e.g., NGC 3504, figure 8 by EPB08). 

\begin{figure*}
\centering
\includegraphics[width=18cm]{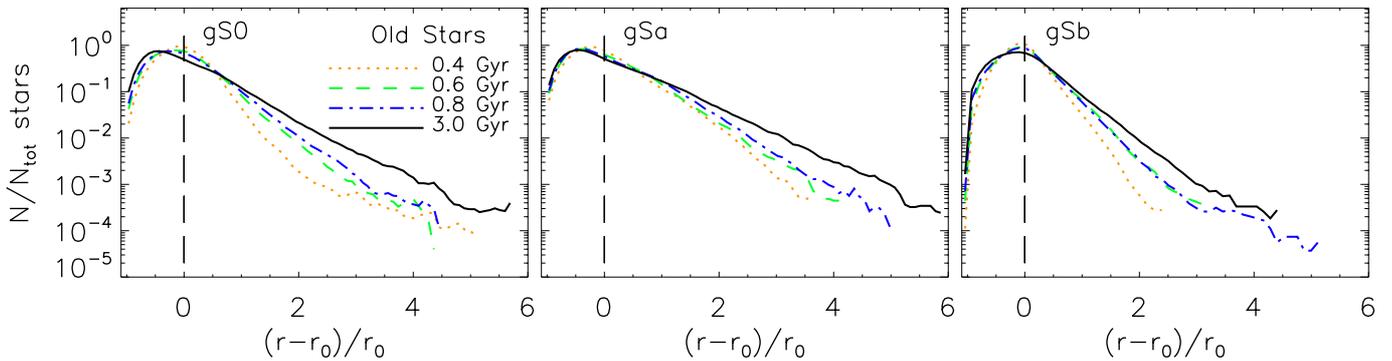}
\caption{
The distribution of fractional changes in initial radius, $(r-r_0)/r_0$, for the same time outputs as in Fig.~\ref{fig:gSall}. Particles out to 40~kpc are considered. The distributions are asymmetric across the zero value (vertical dashed line) indicating that the discs expand. 
\label{fig:rfr_all}
}
\end{figure*}

Inspecting the evolution of the metallicity profile of gS0 (first row, second column of Fig.~\ref{fig:gSall}), we find that the initially strong gradient quickly flattens at $r>h_d$. We observe an upward kink at $\sim15-20$~kpc for early and later times, respectively, similarly to the break seen in the density profile but significantly further out in the disc. At $t=3$~Gyr, the metallicity at $r\sim35$~kpc ($\sim9h_d$) is consistent with that at $\sim7-8$~kpc of the initial disc (red solid curve). This suggests that the stellar population reaching $35$~kpc originates at $\sim7-8$~kpc, which is near the bar's initial OLR and final CR, thus the radius of the strongest migration taking place in the outer disc \citep{mf10}. The occurrence of breaks in the metallicity profiles is not surprising, given the breaks in the surface density and the existence of an initial metallicity gradient in our simulations. 

The galaxy model just described lacks a gaseous component. The second column of Fig.~\ref{fig:gSall} presents the gSa model, which is identical to gS0, except for the 10\% gas present initially. In this case the breaks in the density profiles at different times span a larger range, from $\sim14$ to $\sim25$~kpc for earlier and later times, respectively. Again, very remarkably, the initially truncated disc grows as a single exponential between $\sim1.5h_d$ and the break radius, this time doubling the disc extent by reaching $\sim6h_d$ before it steepens. This model is very much consistent with the Type II.o--CT profile, where the break in the surface-brightness profile is found {\it outside} the bar's OLR (e.g., NGC 2950, figure 8 by EPB08), which has so far been attributed to star-formation thresholds. We discuss this further in Sec.\ref{sec:match}.  

\begin{figure*}
\centering
\includegraphics[width=17cm]{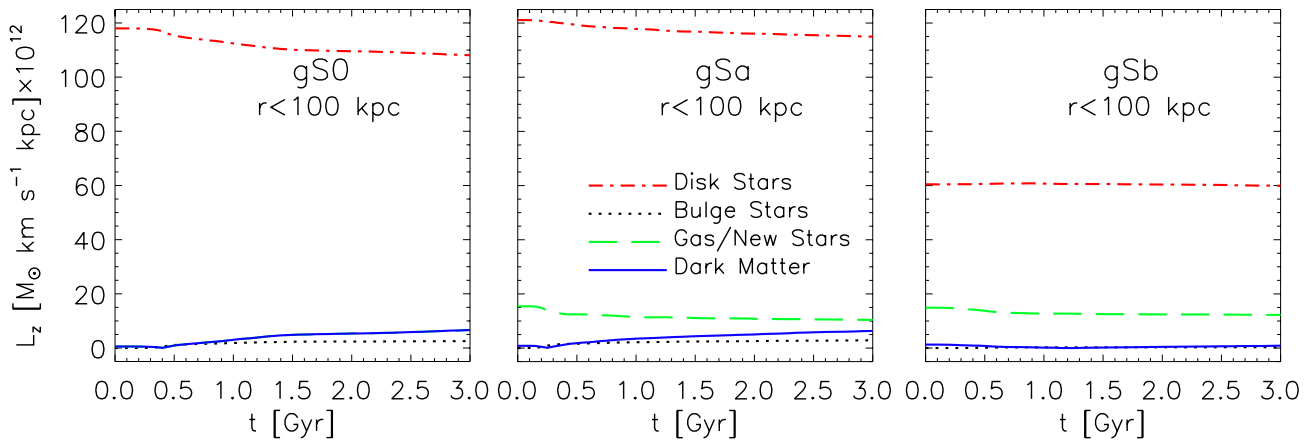}
\caption{
Time evolution of the vertical angular momentum, $L_z(t)$, for different galactic components. Different style and color curves plot the old stellar component (disc only), the bulge, the gaseous disc component, and the dark matter halo, as marked in the figure. From left to right we present the gS0, gSa, and gSb models.   
\label{fig:L_tot}
}
\end{figure*}

The gSa metallicity profile also exhibits an upward kink at later times, but displaced at larger radii, as seen in the density.

The middle column, third row panel shows the radial density profile of the initial gas distribution of the gSa model. At later times, these hybrid particles comprise a mixture of gas and newly formed stars, which is reflected in their dynamics. We see the typical behavior expected of gas dynamics in barred discs, where bar torques and exchange of angular momentum across the bar's CR allow the gas to flow towards the galaxy center (e.g., \citealt{friedli93}), yielding the formation of a ``pseudo-bulge" \citep{kormendy93, courteau96, kormendy04}. Outside the CR, the time evolution of the radial density profile of hybrid particles resembles that of the old stellar component discussed above. A notable difference is the break occurring even further out. This is most likely related to the larger initial scale-length of the gaseous disc (5 instead of 4 kpc).

How do we explain the differences in the evolution of the density and metallicity profiles between the gS0 and gSa models? Compared to the gS0 model, the gSa's bar is similar in length but slightly weaker at all times (Fig.~\ref{fig:om}). However, the gSa's spirals are stronger, especially during the first Gyr (Fig.~\ref{fig:bars}). This is caused by the initial gas component, which keeps the disc cool and thus more unstable. As described by \cite{mf10}, the resonance overlap of bar and spirals results in a very efficient exchange of angular momentum. How effective the migration is depends strongly on the strength of both the bar and spiral structure (see Fig.~4 in \citealt{mf10}). Hence, even though the gSa's bar is weaker, the presence of stronger spirals results in a substantially more massive outer disc.

\subsection{The effect of a weaker bar: the gSb model}

The third column of Fig.~\ref{fig:gSall} shows the gSb model, which has an initial gaseous disc with mass 20\% that of the stellar disc. Although the initial stellar disc is truncated at $\sim14$~kpc (compared to 12~kpc for gS0 and gSa), we observe only weak disc extension in the 3~Gyr time evolution shown in the figure for both old stars and gas. The metallicity profile at later times becomes shallower (occurring at $\sim$~15-17 kpc), as opposed to the upturn in the case of the gS0 and gSa models. The lack of a significant mass transfer outward can be attributed to the much weaker bar and spirals. While comparable in strength to the other models for a short period of time around 300~Myr, the gSb bar decreases in amplitude quickly to a maximum of $\sim0.65A_{2}/A_0$ (Fig.~\ref{fig:om}) and length 4.5~kpc. This is slightly weaker than the gSa bar at later times, but significantly shorter. The spiral structure seen in the $A_{2}/A_0$ Fourier component (bottom row of Fig.~\ref{fig:bars}) is similar to that of the gS0 model. Due to the gSb bar's slowing down at later times, its resonances move $\sim$~2.5~kpc outward in the disc. This ``resonant sweeping" is to be compared to $\sim6$ and $\sim7.5$~kpc in the case of gS0 and gSa, respectively. As in the case of gSa, the gSb density profile evolves to a Type II.o-CT profile, but with the break occurring at the initial truncation radius. 

In all three models discussed above, the initial disc scale-lengths of the middle exponential (between the inner and outer breaks) increase. This is apparent from the flattening of this part of the density profiles, where the scale-length is given by the negative of the slope of a line fit, since the vertical axis is logarithmic. This is in agreement with a number of previous works (e.g., \citealt{foyle08, debattista06}).

\subsection{Redistribution of initial radii}

Another way to illustrate the discs' spreading out is by considering the redistribution or initial stellar radii. In Fig.~\ref{fig:rfr_all} we plot the distribution of the fractional changes in initial radius, $(r-r_0)/r_0$, for the same time outputs as in Fig.~\ref{fig:gSall}. Particles out to 40~kpc are considered. We see that the distributions are strongly asymmetric across the zero value (vertical dashed line), increasingly so for later times, indicating that the discs expand. Increase in stellar radii can be up to 6 times their initial value for gS0 and gSa. On the other hand, negative changes in radius are only about 100\%. The stronger changes for the gS0 and gSa models are consistent with their more extended discs shown in Fig.~\ref{fig:gSall}. 

\section{Angular momentum exchange among different galactic components}
\label{sec:halo}

Before we consider the angular momentum redistribution in the disc itself, we first study its exchange between the different galactic components: disc, bulge, gas, and dark matter halo. Earlier works have related the formation of a bar and its strength to the halo concentration (e.g., \citealt{athanassoula02,bournaud05}), as well as its triaxiality \citep{machado10}. As bars form, the inner disc looses angular momentum since bars live within their CR. Some of this lost angular momentum can be acquired by the dark matter. Further exchange with the halo can slow down bars even more. However, we must note that even in simulations with rigid halos, bars do slow down \citep{combes81}. In that case all lost angular momentum goes into the outer disc. Our simulations have live halos and we thus expect that some disc angular momentum is lost to the halo.  

Following \cite{dimatteo08} (their Fig.~4), we show the time evolution of the total vertical angular momentum, $L_z$, of the different galactic components for the gS0, gSa, and gSb models, in Fig.~\ref{fig:L_tot}. We consider particle distribution within 100~kpc around the galactic center. Different colors and line-styles, shown in the middle panel, indicate the old stellar component (disc stars), the bulge, the gas/new stars, and the dark matter halo. We expect exchange of $L_z$ among all these components. 

For the gS0 model (left panel), the initially non-rotating dark matter halo absorbs about 2/3 of the angular momentum lost by the disc. Remarkably, the remaining 1/3 is acquired by the (also initially non-rotating) bulge. These exchanges are clearly seen to take place after the bar is formed and continue until the end of the simulation, while decreasing once the bar buckles (and thus weakens) at $t\sim1.5$~Gyr (see Fig.~\ref{fig:om}). 

The gSa model shows about half the loss in stellar component $L_z$ compared to gS0. This is consistent with the larger mass transferred into the outer disc we found in the previous Section. However, we note that the gain by the halo and bulge (identical to those of the gS0 model) is very similar to the gS0 case. The additional angular momentum here comes from the gaseous component (dashed green curve), not present in the gS0 simulation. Thus, the gas disc either exchanges $L_z$ directly with the non-rotating components or passes it on to the stellar disc. Therefore, the presence of gas in the disc not only supports stronger spiral structure, but also allows the stellar disc to retain most of its angular momentum, making it available for extending the disc. 

Finally, for the gSb model (right panel of Fig.~\ref{fig:L_tot}) we see that the change in $L_z$ for the stellar disc is consistent with zero. Small variations are seen for the gaseous disc, the bulge and halo. While virtually no disc angular momentum is lost to the halo and bulge, the gSb disc was found to extend the least, which is related to its weak bar.

As seen in Fig.~\ref{fig:L_tot}, only about 10\% and 5\% of the disc angular momentum is lost to the halo and bulge for the gS0 and gSa models, respectively, and effectively none for gSb. This small loss of angular momentum to the halo is not surprising, given that we have maximum discs (see e.g., \citealt{debattista98}).

Since in this work we are only concerned with the angular momentum in the z-direction, we simply refer to it as $L$ for the remainder of the paper.

\section{Disc morphology and angular momentum exchange}
\label{sec:multi}

In Section~\ref{sec:gSall} we have related the extension of discs beyond their initial truncation radii and the occurrence of breaks in their density profiles to the strength of bars and spirals. We now examine the discs' morphology and dynamics in more detail in the effort to determine more precisely the causes for the changes we see in Fig. \ref{fig:gSall}. 

\subsection{Discontinuities in spiral arms}
\label{sec:discon}

\begin{figure*}
\centering
\includegraphics[width=18cm]{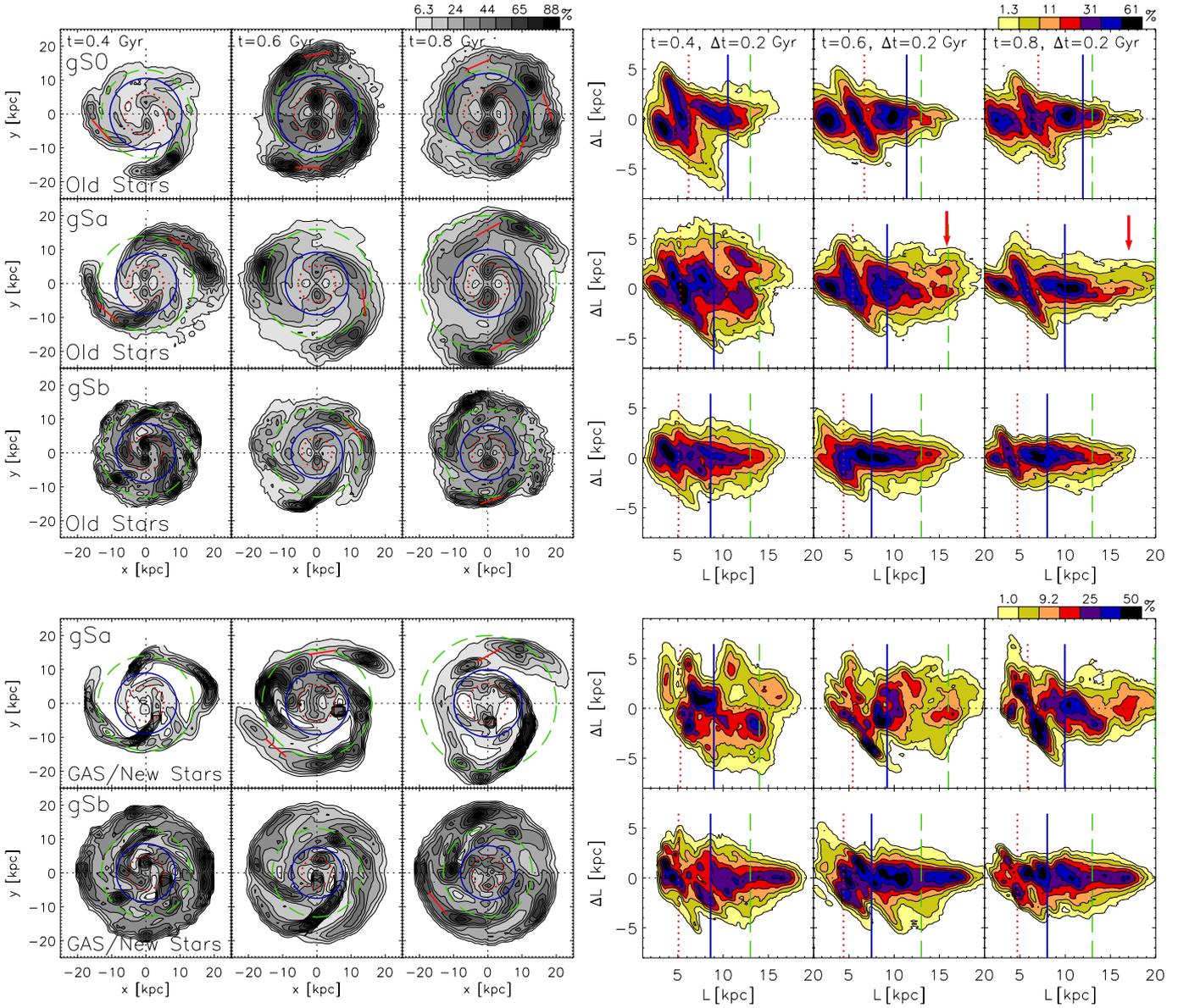}
\caption{
{\bf Columns 1-3:} Differential, face-on, density of the gS0, gSa and gSb discs, for three different times. The first three rows show the ``old" stars, while the last two rows show hybrid particles. All bars are aligned with the vertical axis. Some of the spiral discontinuities are labeled by short, solid, red lines.{\bf Columns 4-6:} Incremental changes in angular momentum, $\Delta L$, in time intervals of 200~Myr, centered on the time of the snapshots shown in the first three columns. Both axes are divided by the estimated rotation curve for each galaxy and the total mass in each radial bin, thus the values displayed are approximately equal to galactic radius in kpc. The bar's CR, its OLR, and the truncation radius are shown as the dotted-red, solid-blue, and dashed green circles (density) or vertical lines ($\Delta L$) in each plot. The red arrow heads indicate strong increases in $\Delta L$ near the disc break for the gSa model, resulting in the more extended disc seen in Fig.\ref{fig:gSall}. 
\label{fig:gSall_am}
}
\end{figure*}

Here we relate the breaks seen in the radial density profiles in Fig.~\ref{fig:gSall} to the discs' morphology. In the left three columns of Fig.~\ref{fig:gSall_am} we show density contours for the gS0, gSa, and gSb models. The first three rows show the old stellar population at different times, while the last two rows present the gas/new stars of the gSa and gSb models. In order to emphasize the asymmetries in the outer disc, the quantity plotted is the differential density, $\Sigma_{dif}$, as done by \cite{mq08}. This is computed as $(\Sigma-\Sigma_{axi})/\Sigma_{axi}$, where $\Sigma$ is the raw value of the stellar density and $\Sigma_{axi}$ is the azimuthally averaged density at each time with radial bin-size of 0.5~kpc. The times of each snapshot are shown in the first row. The bars are aligned with the vertical axis and their CR and OLR radii are shown by the dotted red and solid blue circles, respectively. Note that resonances can shift inward and outward with time if the bar speeds up or slows down. The green dashed circles show the location of the breaks seen in Fig.~\ref{fig:gSall}. For the sake of comparison, for both the stellar and gas discs the green circles show the breaks in the stellar density profiles.

Unlike $\Sigma$ (not shown here, but see Fig.~1 in \citealt{minchev11a}), the differential density $\Sigma_{dif}$ displays strong variations with radius with prominent discontinuities (strong decrease in amplitude) in the spirals. Some of these minima are shown in the Figure by short, solid, red lines bridging the gaps. These gaps in the outer spirals appear to move outward as time progresses for the gSa, but not significantly so for the other two models.  This is very similar to the occurrence of the breaks in the density profiles (green dashed circles in these plots), which we described in Sec.~\ref{sec:gSall}. Interestingly, the breaks always lie just inside the outermost gaps in the spiral arms, where the spiral pitch angles decrease, as is expected when spirals get weaker. 

The discontinuities in the spiral arms outside the bar's OLR move outward with time for gSa, but not significantly so for gS0 and gSb. We showed in Fig.~\ref{fig:gSall} that even at the final time (3~Gyr) the gS0 disc break occurs at 17~kpc, coinciding exactly with the bar's OLR. It, therefore, appears that the slowing down of the gS0's bar (which moves the OLR radius outward) produces the dominant effect, i.e., this galaxy is bar dominated. Conversely, the discontinuities in the case of gSa, which happen at about twice the OLR radius at $t=0.8$~Gyr, indicate stronger instabilities in the outer disc. We study in detail the mode coupling in the gSa disc in Sec.~\ref{sec:coupling}. The strong spirals in this simulation result from the increased surface density in the outer regions due to the diffusion caused by the bar, while the disc is maintained cool by the initial 10\% gas present. Although the gSb model starts with 20\% gas, the disc does not extend significantly. The reason for this is the larger initial gas fraction, which prevents the formation of a strong bar due to the central mass concentration resulting from the gas inflow toward the center as the bar forms. Additionally, \cite{mf10} demonstrated that the mixing by resonance overlap of bar and spiral structure is a non-linear mechanism, strongly dependent on the strength of the perturbers. Therefore, the smaller bar, which forms during the gSb simulation cannot transfer enough mass in the outer disc to make it unstable enough to sustain spiral instabilities beyond the initial truncation radius. Another way of looking at this is comparing the amount of stellar mass redistributed for each model: inspecting the area inclosed between the red and black curves, we find a big difference in favor of the gS0 and gSa models compared to gSb (relatively speaking since the gSb disc has a smaller total mass). In turn, due to conservation of angular momentum for the entire disc (no significant loss of angular momentum to the live halo was found in Section~\ref{sec:halo}), we expect less mass transfer outwards for gSb, and, thus, less disc extension.

Discontinuities in spiral arms can indicate changes in the dominant pattern (transition between inner and outer structure) or constructive/destructive interference between different spiral modes (different waves propagating at the same radius). Multiple patterns are expected in a resonance-coupled system, as recently demonstrated by \cite{quillen11}, where in an N-body simulation both the inner and outer spirals waves were found to be coupled with the bar. For example, the lack of rotational symmetry in many of the snapshots shown in Fig.~\ref{fig:gSall_am} is indicative of an $m=1$ mode, which may be resulting from interference between two patterns of different multiplicity (e.g., a four- and a three-armed spiral waves). We show that this is indeed the case in Sec.~\ref{sec:coupling}. Observationally, lopsidedness has been found to be a common property of spiral galaxies. For example, \cite{zaritsky97} have shown that $\sim$~30\% of field spiral galaxies in a magnitude-limited sample exhibit significant lopsidedness in the outer discs. Even classical Galaxy analogues, such as NGC~891 exhibit big asymmetries, clearly resolved as gas accreting along a large filament or arm \citep{mapelli08}.

The first three panels of the bottom two rows in Fig.~\ref{fig:gSall_am} show density contours of the gas/new stars component for the gSa and gSb models. They appear quite similar to the stellar counterparts we just described. The colder population in these plots makes it clearer that gaps in the spiral arms indicate the transition to a different set of structure (possibly with different pattern speed). For example, for the gSa, the arm at $(t,x,y)=(0.8,10,15)$ is now clearly seen to be independent of the inner nearby one. Two additional prominent outer arms as seen at the earlier time as well: compare the gaseous and stellar discs at $t=0.6$~Gyr. The case is similar for gSb. 

\subsection{Angular momentum exchange at different radii}
\label{sec:am}

We now study the angular momentum exchange at different radii in our galactic discs. In Fig.~2 by \cite{minchev11a} we presented the changes in angular momentum as a function of radius for the duration of the simulations (3~Gyr) for gS0, gSa, and gSb. In contrast, in columns 4-6 of Fig.~\ref{fig:gSall_am} we show the {\it incremental} changes in the time intervals 0.3-0.5, 0.5-0.7, and 0.7-0.9~Gyr. We estimate these as $\Delta L(r)=L_1(r)-L(r)$, where $L(r)$ and $L_1(r)$ are the ``initial" (i.e., t=0.3, 0.5, and 0.7) and ``final" (i.e., t=0.5, 0.7, and 0.9) angular momenta as function of radius at each time step. Thus, each plot shows how much migration has occurred as a function of radius during the time step of 200~Myr, centered on the time of the snapshots shown in the first three columns of the same figure. To properly display density contours we consider only stars with $r>2$~kpc. The bar's CR and OLR are shown as the dotted red and solid blue vertical lines in each plot. Note that in all cases resonances shift outward due to the bar's slowing down, not very significantly for the gSb model. The estimated angular momentum at each radial bin is divided by the rotation curve appropriate for each galaxy model and the total mass at that bin. Therefore, the units of $L$ and $\Delta L$ are kpc. This means that, for example, the group of stars at ($L,\Delta L)\approx(3,5)$ is shifted radially outwards by $\sim5$~kpc. Similarly, stars losing angular momentum have their mean (guiding) radii shifted inwards. 

In all three models the regions around the bars' CR radii (red dotted vertical lines) display strong exchange of angular momentum. The amount of mixing across the CR is approximately constant for all times, while the outer disc regions display structure which varies with time. As we will show later, this is related to the interplay of spiral waves of different multiplicity. The amount of migration taking place outside the bar's CR for each simulation is consistent with the discs' extension shown in Fig.~\ref{fig:gSall}: the gSa disc is affected the most, especially in the outer parts. This is reflected in the fast radial advancement of the disc break (dashed green vertical line) compared to the other models. The red arrow heads indicate strong increases in $\Delta L$ for the gSa model, giving rise to the disc extension. Such peaks are not seen for the other two models. The constant change in structure in the $L-\Delta L$ plane for the gSa disc implies that it is not a single perturber that drives mass to the outer disc but, rather, a network of waves.

It is clear from these plots that to drive the break outwards, the disc needs to be unstable in the region near the break. In our simulations (not considering gas accretion), this is possible if enough mass is transferred outward, while keeping the disc cool so that the Toomre instability parameter, $Q_t\sim\sigma_r/\Sigma$, is sufficiently low. Given the strong bar and spiral structure in the gSa model, and the presence of the initial gas fraction, it is natural that this disc extends the most. The larger area at positive $\Delta L=0$ in the gSa outer disc indicates that stars suffer a large increase in their home radii (on the order of 3~kpc). The angular momentum exchange here is balanced approximately by the smaller scale inward migration ($\sim1$~kpc, red and orange contours) for a larger mass fraction. This is in contrast to the gS0 and gSb discs, where changes above and below the $\Delta L=0$ are approximately equal. 

\begin{figure}
\includegraphics[width=9cm]{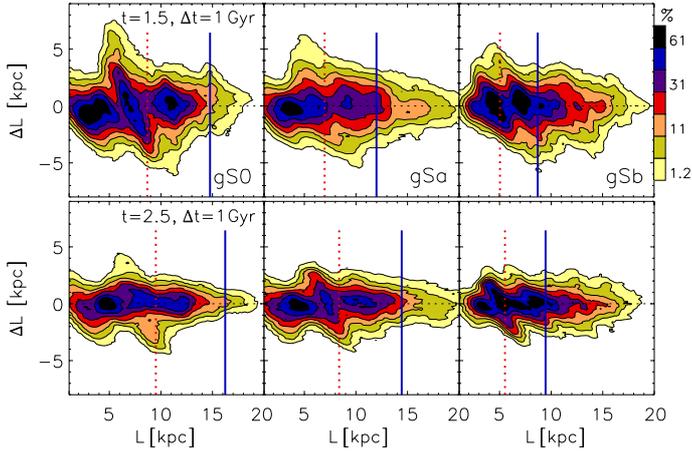}
\caption{
Changes in angular momentum in the time periods $t=1-2$~Gyr (top) and $t=2-3$~Gyr (bottom) for the gS0, gSa, and gSb models. The initial angular momentum $L$ here corresponds to $t=1$ and $t=2$~Gyr, respectively. The time averaged value of the bar's CR and 2:1 OLR are shown by the dotted-red and solid-blue lines. It is evident that the bars are the most effective drivers of radial migration through their CR, despite the fact that they are not transient, but only slowly evolving. This is true for the entirety of the simulations, as this figure shows.
\label{fig:3gyr}
}
\end{figure}

The bottom two rows of Fig.~\ref{fig:gSall_am} are similar to the above plots, but display the gas/new stars component for the gSa and gSb models. As in the old stellar population, here we also find prominent increases in $\Delta L$; these are even more strongly emphasized since circular orbits are affected more by the perturbations \citep{mnq07}.

To see whether the bar is an effective driver of radial migration for times grater than 1~Gyr, in Fig.~\ref{fig:3gyr} we show the changes in angular momentum in the time period $t=1-2$ (top) and $t=2-3$~Gyr (bottom) for the gS0, gSa, and gSb models. The initial angular momentum $L$ here corresponds to $t=1$ and $t=2$~Gyr, respectively. The time averaged value of the bar's CR and 2:1 OLR are shown by the dotted-red and solid-blue lines. We see that the effect of the bars' CR is evident throughout the entire simulations. Even for the gSb model (which hosts the smallest bar) changes due to the bar are similar to those of an outer spiral between the bar's CR and the OLR. This fact renders the bars in our simulation the most effective drivers of radial migration, despite the fact that they are not transient, but only slowly evolving. This is surprising in view of the theory developed by \cite{sellwood02}, where the CR is effective only if structure is transient. This is consistent with the findings of \cite{brunetti11}.

To make a better judgement about the drivers of radial migration in our discs, we now estimate the pattern speeds.

\section{Pattern speeds}
\label{sec:power}

We construct power spectrograms using the procedure outlined by \cite{quillen11}, Sec.~2.3. During an N-body simulation with a live halo (as in our simulations), lopsided modes may develop and the bulge and central disc of the galaxy may not remain at a fixed position. We, therefore, subtract the position of the centroid of the galaxy bulge prior to computing the Fourier coefficients. A sufficient number of time outputs in a given time window is required to compute a spectrogram. Since structure may be transient and pattern speeds may vary with time, a small time-step between outputs is needed for an accurate determination of the temporal evolution of density waves in a galactic disc. For the simulations presented here we have time outputs every 10~Myr.  

\begin{figure*}
\centering
\includegraphics[width=18cm]{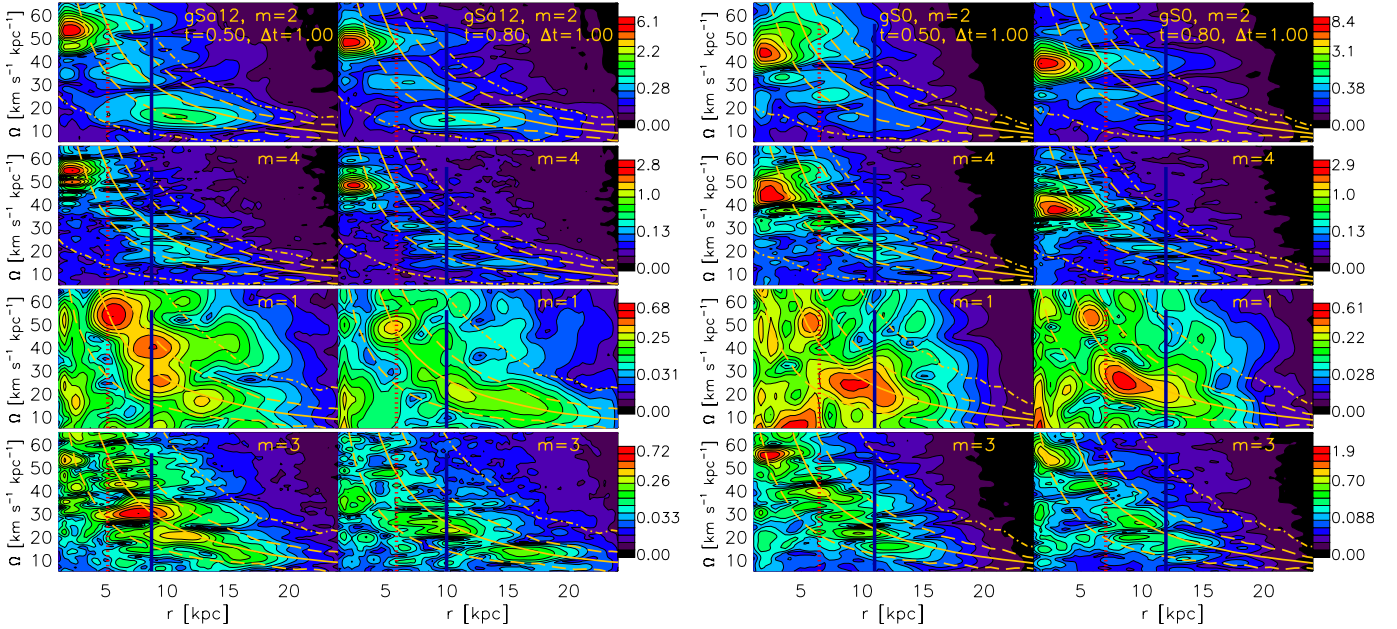}
\caption{
Power spectrograms (from top to bottom) of the $m=2, 4, 1$ and 3 Fourier components for the gSa (columns 1-2) and gS0 (columns 3-4) models. Spectrograms are computed in a time window $\Delta t=1$~Gyr centered on $t=0.5$ and $t=0.8$~Gyr. For $m=2$ and $m=4$, the resonance loci are plotted for the CR (solid), 4:1 LR (dashed) and 2:1 LR (dot-dashed) orange lines, computed as $\Omega$, $\Omega\pm\kappa/4$, and $\Omega\pm\kappa/2$, respectively. For the $m=1$ and $m=3$ cases, we plot $\Omega$, $\Omega\pm\kappa$, and $\Omega\pm\kappa/3$. Both bars and spirals are seen to slow down at later times. The dotter-red and solid-blue vertical lines show the radial location of the bar's CR and OLR.
\label{fig:gSa0_omega}
}
\end{figure*}

In Fig.~\ref{fig:gSa0_omega} we show spectrograms constructed from the $m=2, 4, 1,$ and 3 (top to bottom) Fourier components for the gS0 (columns 1-2) and gSa (columns 3-4) models. For each galaxy, spectrograms are computed at $t=0.5$ and $t=0.8$~Gyr with a time window $\Delta t=1$~Gyr, as indicated in each panel. The vertical axes show the pattern speed in units of km/s/kpc and the horizontal axes plot the galactic radius in kpc. For $m=2$ and $m=4$, the resonance loci are plotted for the CR (solid), 4:1 LR (dashed) and 2:1 LR (dot-dashed) orange lines, computed as $m\Omega$, $m\Omega\pm\kappa/4$, and $m\Omega\pm\kappa/2$, respectively. For the $m=1$ and $m=3$ cases, we plot $m\Omega$, $m\Omega\pm\kappa$, and $m\Omega\pm\kappa/3$. The color contours represent the strength (in arbitrary units) of the features indicated in the color bar. 

Firstly, we note that the waves we see in the spectrograms may not relate directly to the morphology of our galactic discs. What appeared as well defined spiral arms in the stellar density plots of Fig.~\ref{fig:gSall_am} may be structures comprised of superposed waves moving at different angular velocities. This will become apparent later in this Section.

The structure seen in the $m=2$ power spectra (top row of Fig.~\ref{fig:gSa0_omega}) indicates bi-symmetric waves, such as a central bar and two-armed spirals. For example, the gSa bar is identifiable by the strongest, fast horizontal feature in the inner disc (between 0 and 6 kpc). Two additional features (spirals) are present in the outer disc at lower pattern speeds. The strongest, slower spiral has its 2:1 ILR near the location of the bar's CR, as expected in the case of mode coupling (e.g., \citealt{tagger87,sygnet88}). By comparing the spectrograms at $t=0.5$ and $t=0.8$~Gyr, we see that both the gSa and gS0 bars slows down at the later time, in agreement with Fig.~\ref{fig:om}. While it is easy to discern two dominant spiral patterns for both the earlier and later times for gSa, structure in the outer disc is not very clear in the later gS0 plot. This is related to the weak spirals of this model at times $\gtrsim0.5$~Gyr, as seen in the bottom panels of Fig.~\ref{fig:bars}. Note that the $m=2$ spirals in all plots end close to the CR (solid curves). This fact, as well as the common decrease in pattern speed for both bar and spirals is consistent with the findings of \cite{quillen11}, where a dissipationless simulation was studied. The authors interpreted this as a sign that the system was coupled.

The $m=4$ power spectra are shown in the second row of Fig.~\ref{fig:gSa0_omega}. These represent four-armed spirals, as well as the $m=4$ symmetry in the bar (its first harmonic) at $2\Omega_{bar}$. Due to the factor of two higher frequency, the bar pattern speed is much better defined in this plot. Given that the bar is mostly bi-symmetric, its strength here is roughly half of the $m=2$ component. 

The $m=3$ Fourier component is shown in the bottom row of Fig.~\ref{fig:gSa0_omega}. Two dominant (inner and outer) three-armed spiral waves are found to propagate in the gSa disc, in addition to the $m=2$ and 4 structure just described. These are weaker by about a factor of 0.5 compared to the amplitudes of the outer, two-armed spiral wave, as evident by comparing the corresponding color bars. Three-armed patterns in the gS0 model are also present, but not well defined and weaker, as is the $m=2$ structure, due to the lack of gas and consequently hotter disc.

We finally discuss the third panels of Fig.~\ref{fig:gSa0_omega}. Here we show the $m=1$ Fourier component, which represents lopsided structure, such as one-armed spirals. We note that the position of the centroid of the inner disc has been subtracted, therefore, the features seen in these plots are most likely due to the interaction of different waves at certain radii. It is remarkable that the strongest $m=1$ features in both models (better visible in gSa) coincide with the $m=3$ spirals in both radial extent and power (see the color bars). The inner/faster and outer/slower $m=1$ and $m=3$ waves start at the bar's CR and OLR, respectively, hinting that these modes may be coupled. \cite{brunetti11} also found an $m=1$ pattern in an N-body barred disc and suggested that this feature can be a strong driver of stellar diffusion (migration); the nature of this perturbation may be similar to what we find here.

In the power spectrograms plotted in Fig.~\ref{fig:gSa0_omega} we found some well defined features, e.g., inner and outer spiral features in the gSa disc. Such a low number of coherent structure in spectrograms averaged over a long time period could mean two things: (i) spirals are recurrent, but form at the same radii and pattern speeds or (ii) spirals are long-lived. In the following section we investigate the longevity of spiral structure in our simulations.

\subsection{Spiral structure longevity}
\label{sec:longevity} 

At present the nature of galactic disc spiral structure is not well understood. Though it is generally accepted that spirals are density waves there exist two competing theories:
(i) transient/recurrent spirals, and
(ii) long-lived, steady-state spirals.

Recurrent spiral instabilities have been reported by \cite{sellwood84} and \cite{sellwood89} in their simulations of self gravitating discs. It was argued by \cite{toomre91} that these transient spiral density waves are due to the swing-amplification mechanism as first formulated by \cite{toomre81}.

\begin{figure*}
\centering
\includegraphics[width=18cm]{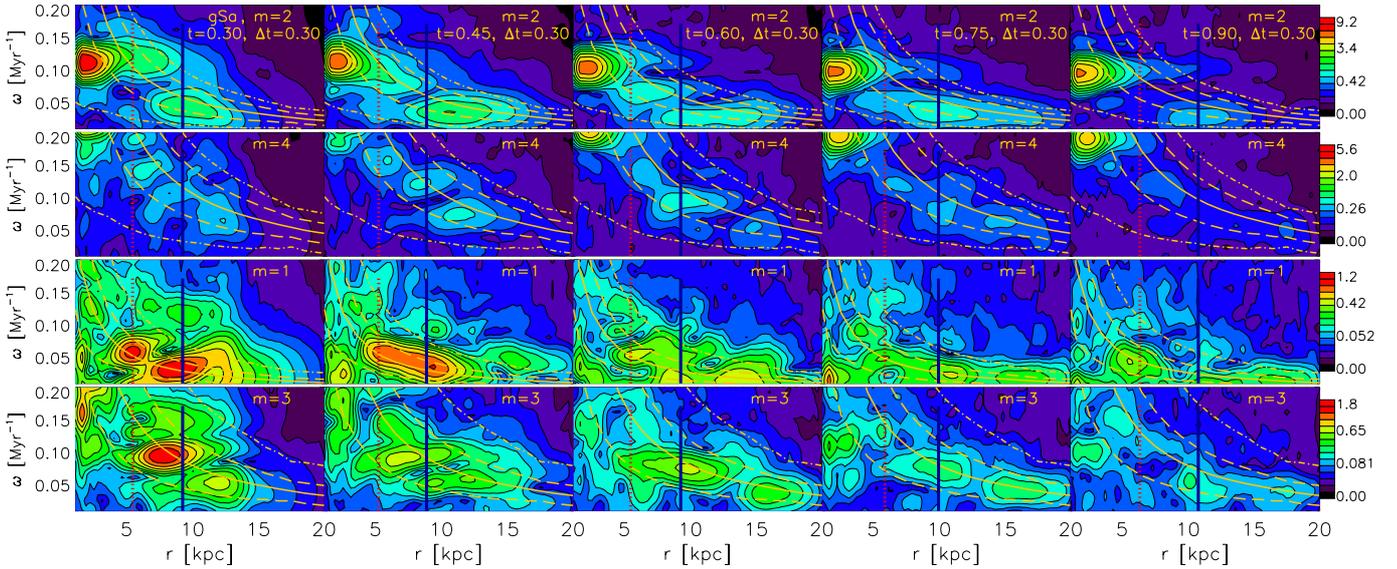}
\caption{
Time evolution of the $m=2, 4, 1,$ and 3 (top to bottom) power spectra for the gSa model. The vertical axis shows the frequency, $\omega=m\Omega$ (in units of km~s$^{-1}$~kpc$^{-1}$), rather than the pattern speed $\Omega$, as done in Fig.~\ref{fig:gSa0_omega}, in order to facilitate a discussion of coupling between different modes. Time outputs are every 150~Myr with a time window $\Delta t=300$~Myr. Contour levels are normalized for each multiplicity to better follow the changes in amplitude. The orange curves show the resonant loci as in Fig.~\ref{fig:gSa0_omega}. It is clear that the outer $m=2$ wave has a stable pattern speed (slowly decreasing as the bar slows down) for over $600~Myr$, while slowly weakening and extending outward with time. In contrast, the inner $m=2$ wave bounces between the bar and the outer one at the beat frequency of the latter two patterns. The $m=3$ waves are also present for the entire time, shifting outwards with time and largely coinciding in radius with the $m=1$ features. 
\label{fig:gSa_omega2}
}
\end{figure*}

Alternatively, the concept of quasi-stationary density waves was developed (mostly analitically) by \cite{lin69} and culminated in the work by \cite{bertin89a,bertin89b} and \cite{lowe94}. While thought to always produce short-lived spirals, N-body simulations have been constructed to yield long-lived spiral density waves lasting for over five rotation periods, by introducing an inner Q-barrier to shield the 2:1 ILR \citep{thomasson90,elmegreen93,donner94,zhang96}. We note that the recent study by \cite{sellwood11} have questioned the results of these works, claiming that spiral pattern speeds have not been measured properly since multiple modes were taken for a single one. 
 
We want to find out how long-lived are the spirals in our simulations. As seen in the above introduction, estimating spirals pattern speeds has been proven to be a hard task. How do we ensure we do this properly? Measuring the time evolution of the spiral structure seen in the stellar or gas density is not appropriate since multiple waves (if present, as in our simulations) can interfere and thus lead to the wrong result. Power spectrograms, on the other hand, allow for estimating the frequencies of the actual waves (at different multiplicities) and is thus, perhaps, the most appropriate way.

In Fig.~\ref{fig:gSa0_omega} we plotted Fourier power spectrograms, computed using a time window of 600~Myr. Short transient spiral waves cannot be identified in this plot. To look for spiral lifetimes we need to decrease the time window used, as well as the time spacing. Therefore, we construct spectrograms in a time window of 300~Myr, i.e., half of the one we used before, and plot these every 150~Myr from $t=150$ to $t=900$~Myr. We chose to examine the gSa disc because of the better signal to noise ratio, given its strong spirals. We present our results in Fig.~\ref{fig:gSa_omega2}. In contrast to Fig.~\ref{fig:gSa0_omega}, the vertical axes here show the angular frequency, $\omega$, rather than the pattern speed $\Omega=\omega/m$, in order to facilitate a discussion of coupling between different modes (next Section). Because of the additional factor $m$ here, the bar's $m=4$ response appears at twice the $m=2$ $\omega$-value (visible at later times as the bar slows down). The color bars are normalized for each multiplicity to better follow the variation in amplitudes. 

We first examine the strongest spiral mode, $m=2$, shown in the first row of Fig.~\ref{fig:gSa_omega2}. Despite the smaller time window used, structure still appears coherent. At $t=300$ (first column) a wave of frequency $\omega\sim0.05~{\rm Myr^{-1}}$ forms, similar to the outer feature we saw in Fig.~\ref{fig:gSa0_omega}, first column, extending between the bar's CR and the disc break (see Fig. \ref{fig:gSall}). At later times this wave is always present, increasing in length up to $r\sim18$~kpc at $t=750$~Myr. When animated, this two-armed feature is aways present in the spectrograms, smoothly changing from one time output in the figure to the next. We, therefore, conclude that this pattern has a lifetime $\gtrsim600$~Myr. The smooth decrease in strength, as well as the radial extent with time, is also in agreement with the conclusion that we see the same pattern in all snapshots. In contrast, the inner $m=2$ spiral, while appearing similar to the outer one in Fig.~\ref{fig:gSa0_omega}, we now find it to exhibit strong variations with time. When animated, this feature is seen to be driven by the interaction of the bar with the outer $m=2$ pattern, bouncing back and forth between the two on a timescale consistent with their beat frequency. In other words, every time the (faster) bar encounters the (slower) outer spiral wave, this inner wave is regenerated, speeding up to catch up with the bar ($t=450$ and $t=600$~Myr) and later on slowing down to reconnect with the outer spiral ($t=750$~Myr). Such an inner structure, connecting the bar with the dominant spiral has been reported before and has been proposed to provide an explanation for the nature of the ``long" bar in our Galaxy \citep{athanassoula05, martinez11, romero11, athanassoula12}. An outer pattern with a roughly constant pattern speed has also been reported from simulations of non-barred galactic discs \citep{roskar11, sellwood11}. \cite{roskar11} also saw an inner/faster spiral wave with a strong variation in angular velocity (their Fig.~5). 

We estimated above that the lifetime of the two-armed outer spiral wave in our gSa model is $\gtrsim600$~Myr. At its average rotational frequency of $\omega\sim0.04~{\rm Myr^{-1}}$, this corresponds to $\gtrsim4$ rotations, which is relatively long-lived and comparable to estimates found in previous work (e.g., \citealt{thomasson90,elmegreen93,donner94,zhang96}). It should be noted that the model we consider here has a substantial bar, which may be related to the longevity of spirals, especially if mode coupling is present (see next Section). 

When animated, the $m=1, 3,$ and 4 features are seen to evolve fairly quickly at the earlier times of the simulation and change position and frequency together, sweeping their resonances through the disc. At later times these waves become longer-lived: the $m=1, 3$ and 4 features at $r\approx8$ and 16~kpc, seen in the fourth column of Fig.~\ref{fig:gSa_omega2}, form at $t\approx50$~Myr and persist until the final time shown in Fig.~\ref{fig:gSa_omega2}, while slowing down and moving to larger radii. The coincidence in radial positions of all these waves at different times of the disc evolution strongly suggest that these modes are coupled. In Section~\ref{sec:coupling} we investigate this possibility.

\subsection{Relationship between density waves and spiral arms}
\label{sec:den_sp}

Here we compare the spiral waves found in the power spectrograms to the spiral arms seen in the stellar density plots. We turn our attention to Fig.~\ref{fig:gSa_omega3}, where the top four rows are the same as Fig.~\ref{fig:gSa_omega2} and the fifth row shows differential stellar density contours of galactic azimuth versus radius, as was done by\cite{quillen11}. Here we see the two halves of the bar as the horizontal strong features at low radii and $\phi=90^\circ$ and $270^\circ$, while spiral structure appears as lines of negative slope. Rotation is in the positive vertical direction. The corresponding disc face-on contours are shown in the bottom row of the same figure. 

An attempt is made to match the waves of different multiplicities to the spiral features seen in the stellar density. Pink vertical lines connect the ends of the most likely matches between waves and material arms. For most times shown, we can identify the slow, two-armed wave and some outer $m=3$ and $m=1$ modes. Although for most time outputs the face-on stellar structure looks two-armed, we can now see that this is only true for the $m=2$ waves up to the discontinuities just outside the bar's 2:1 OLR (blue circle). The outer extensions of the material spirals are, in fact, $m=3$ and $m=1$ modes, easily seen for $t=600$, 750, and 900~Myr. However, a large number of waves cannot be identified in the morphology plots because they interfere. Therefore, one must be careful when making conclusions about properties of spiral structure (such as pattern speed and longevity) based on morphology alone.

\section{Non-linear coupling}
\label{sec:coupling}

Multiple patterns in N-body simulations have been known to exist since the work of \cite{sellwood85} and \cite{sellwood88}, who found that a bar can coexist with a spiral pattern moving at a much lower angular velocity. \cite{tagger87} and \cite{sygnet88} explained this as the non-linear mode coupling between the bar and the spiral wave. These findings were later confirmed by the numerical studies of \cite{masset97} and \cite{rautiainen99}.

According to the theoretical work by \cite{tagger87} and \cite{sygnet88}, two patterns can couple non-linearly as they overlap over a radial range, which coincides both with the CR of the inner one and the ILR of the outer one. This coincidence of resonances results in efficient exchange of energy and angular momentum between the two patterns. The coupling between the two patterns generates beat waves (as we describe below), also found to have LRs at the interaction radii, resulting in a strong non-linear effect even at relatively small amplitudes. \cite{rautiainen99} showed that coupling between a CR and 4:1 ILR, as well as ILR and OLR is also possible in N-body simulations. Waves couple with a selection of frequencies which optimizes the coupling efficiency. Strong exchange of energy and angular momentum is then possible among the coupled waves. 

A mode with an azimuthal wave number and frequency $m_1$ and $\omega_1$, respectively, can couple to another wave with $m_2$ and $\omega_2$ to produce a third one at a beat frequency. The selection rules are
\begin{equation}\label{eq:m}
m = m_1\pm m_2
\end{equation}
and
\begin{equation}\label{eq:om}
\omega = \omega_1\pm \omega_2
\end{equation}
We next look for relation between the different waves found in our gSa model. 

\subsection{Relation between $m=2$ and $m=4$}
\label{sec:m24}

Following \cite{masset97}, we first search for coupling between the bar and the slow $m=2$ outer spiral wave found in our gSa model. Inspecting Fig.~\ref{fig:gSa_omega2}, we find that the slower $m=2$ feature discussed in Sec.~\ref{sec:longevity} has its 2:1 ILR coinciding with the bar's CR at all times. This is in agreement with the expectation that this strong wave is non-linearly driven by the bar. In such a case, it is also expected that this would result in the generation of two additional waves propagating at the beat frequency of the partner waves (if they were allowed). Using Eqs.~\ref{eq:m} and \ref{eq:om} and considering the second column of Fig.~\ref{fig:gSa_omega2}, we add the corresponding frequencies and wave numbers: $\omega_{m4}=\omega_{bar}+\omega_{sp}\approx0.115+0.035=0.15$ and $\omega_{m0}=\omega_{bar}-\omega_{sp}\approx0.08~{\rm Myr^{-1}}$. While we do not show here $m=0$ spectra (ring-like structures in the disc), we can search for the predicted $m=4$ wave. Such a clump, with $\omega\approx0.155$ is identifiable in the $m=4$ spectrogram at the same time output ($t=450$~Myr), in excellent agreement with our estimate above. This feature, found with a much better defined pattern speed in the longer time-window spectrogram of Fig.~\ref{fig:gSa0_omega} (left), propagates between its 4:1 ILR and its CR, where it is then strongly attenuated. Its 4:1 ILR  coincides with the region of resonance overlap between the parent waves in accordance with the expectations \citep{masset97}.  

It is interesting to note that the slow $m=2$ spiral wave strongly decreases in amplitude near its 2:1 ILR at certain times. This is due to the periodic interaction with the inner $m=2$ spiral described in Sec.~\ref{sec:longevity}. This feature reconnects the slow $m=2$ wave and bar every time their phases coincide. Therefore, the $m=4$ wave regrows on the same timescale. 

\begin{figure*}
\centering
\includegraphics[width=18cm]{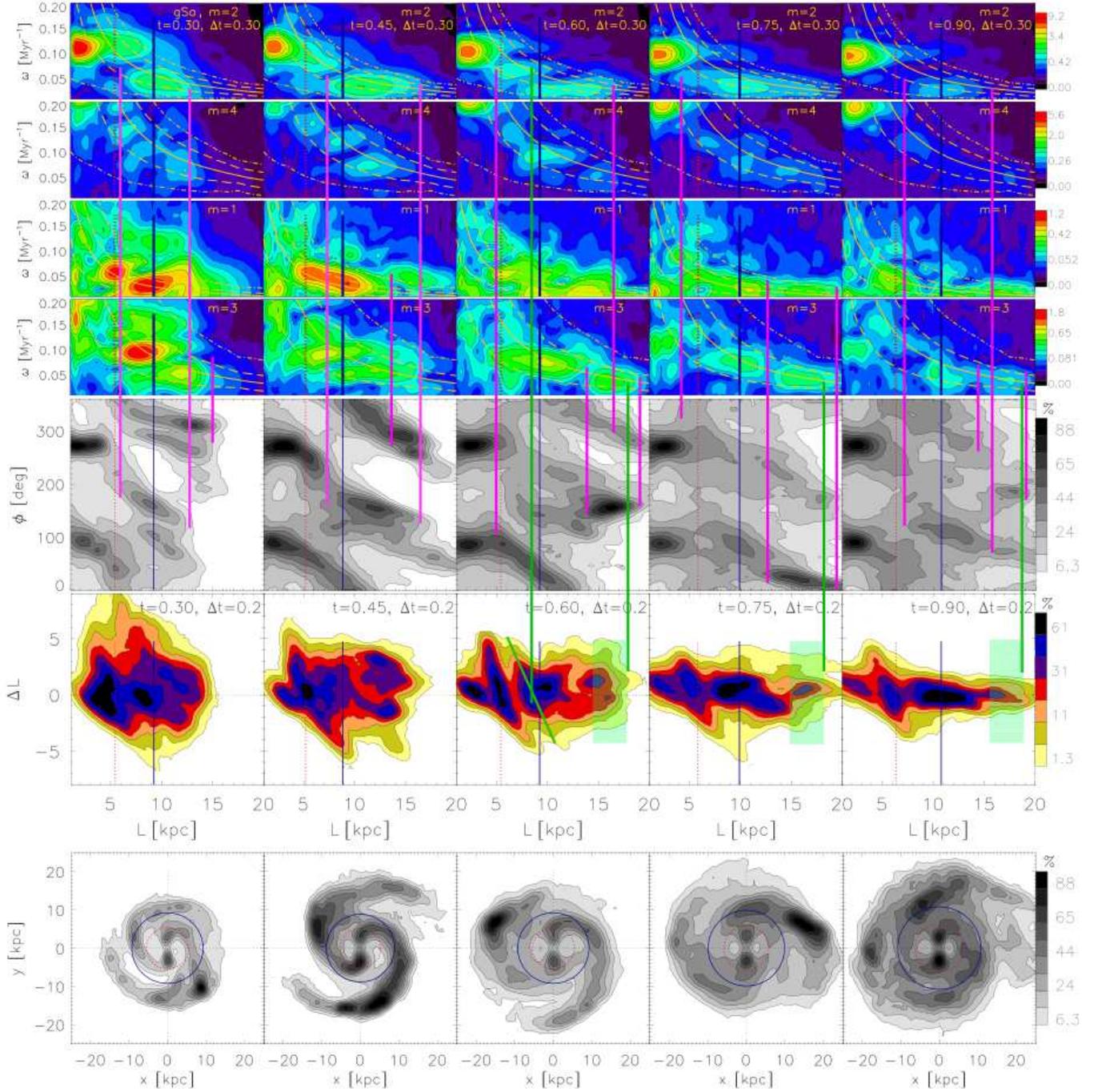}
\caption{
{\bf Rows 1-4} Same as Fig.~\ref{fig:gSa_omega2}. {\bf Row 5:} stellar density contours of galactic azimuth vs radius. An attempt is made to match the waves of different multiplicities to the spiral features seen in the stellar density: pink vertical lines connect the most likely matches. Many waves cannot be seen in the stellar disc due to constructive/destructive interference. {\bf Row 6:} Incremental changes in angular momentum computed in 200~Myr spans, centered on the same times as the spectrograms above. The vertical green lines connect the likely waves/resonances and their effect on $\Delta L$. The effect of the inner $m=2$ spiral, as it sweeps its CR between the bar and the outer $m=2$ wave, is shown by the green slanted line in the middle column. Resonant widths increase as the resonance curves (orange) flatten in the outer disc: clumps in $\Delta L$ are highlighted by the green shaded rectangles in the last three columns. {\bf Row 7:} Face-on differential stellar density contours for the same times.  
\label{fig:gSa_omega3}
}
\end{figure*}

\subsection{Relation among $m=1, 2$ and 3}

Previous works have suggested coupling between $m=1$ and $m=2$ modes \citep{tagger91}, as well as among $m=1, 2$ and 3 modes for both gaseous \citep{laughlin96} and stellar \citep{miller92, quillen11} discs. We now turn our attention to the $m=3$ spectra, shown in the fourth row of Fig.~\ref{fig:gSa_omega2}. For most time outputs shown, we see two major clumps extending between the 3:1 ILR and 3:1 OLR (orange dashed curves). Considering the $t=750$~Myr output (fourth row of Fig.~\ref{fig:gSa_omega2}) for example, we can identify possible coupling among the slower $m=2$ spiral and the $m=1$ and 3 modes centered close to the bar's OLR (blue vertical line): $\omega_{m1}=\omega_{m3}-\omega_{m2}\approx0.07-0.04=0.03~{\rm Myr^{-1}}$. This is slightly higher than the frequency of the $m=1$ mode found just outside the bar's OLR at the same time output ($\omega_{m1}\approx0.02$).

\subsection{Relation among $m=1, 3$ and 4}

With the inner disc's center of mass subtracted, all $m=1$ modes seen in the third row of Fig.~\ref{fig:gSa_omega2} should be related to coupling among patterns with multiplicities $m=n$ and $m=n+1$, where $n$ is an integer. The extent of these features to large radii also speaks in support to this claim. By inspecting spectrograms of different $m$-values, we indeed find such evidence. For example, considering features overlapping in radius in the spectrograms at $t=600$~Myr, we can predict the location of $m=1$ beat waves by subtracting the wave numbers and frequencies of the $m=4$ and $m=3$ features at the same time output. For the waves centered on the bar's OLR (blue vertical line) and $r\approx15$~kpc, respectively, this results in $\omega_{m1} = \omega_{m4}-\omega_{m3} \approx 0.095-0.075=0.02~{\rm Myr^{-1}}$ and $\omega_{m1} \approx0.05-0.03 = 0.02~{\rm Myr^{-1}}$. These $m=1$ waves are clearly seen at the predicted radial positions and frequencies. In addition, a faster $m=1$ mode is found in the same plot very close to the bar's CR moving at $\omega=0.05~{\rm Myr^{-1}}$. This one can be explained as the interaction between the weaker $m=4$ feature with 4:1 ILR near the bar's CR and the strongest $m=3$ wave. Note the large number of resonance overlaps in the bar's CR region: the bar's CR, the $m=4$ ILR, the $m=3$ ILR, and the $m=1$ CR.

\begin{figure}	
\includegraphics[width=9cm]{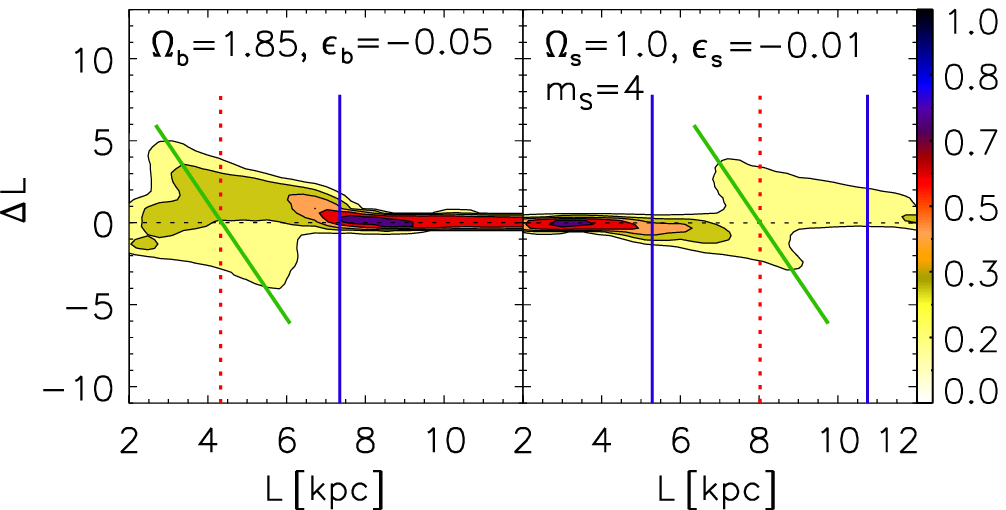}
\caption{
The $L-\Delta L$ plane for a single perturber: a bar (left panel) and a four-armed spiral wave (right panel), from the test-particle simulations presented in \cite{mf10}. The bar and spiral pattern speeds and strengths are indicated by $\Omega_{b,s}$ and $\epsilon_{b,s}$, respectively. The dotted-red vertical lines indicate the CR of each perturber and the solid-blue ones -- the ILR and OLR. The color bar shows the normalized particle density. Such a shape of the distribution (indicated by the green line), with an increase in $\Delta L$ inside and a decrease outside the CR is characteristic for a single perturber, but not when perturbers interact as discussed in the text.  
\label{fig:mf10}
}
\end{figure}

\subsection{Chains of non-linearly coupled waves}

It was already recognized by \cite{lb72} that trailing spiral structure carries angular momentum outwards. Considering a single mode, a wave would absorb angular momentum at its ILR and emit it at the CR or the OLR. We may relate this to the large disc extension seen in our gSa model. In the spectrograms shown in Fig.~\ref{fig:gSa_omega2} we can identify a continuously larger number of waves with increasing multiplicity. These patterns seem to line up in a specific manner.

As discussed in Section~\ref{sec:m24}, the strong, slow $m=2$ spiral feature has its 2:1 ILR at the bar's CR at all times, suggesting non-linear mode coupling between the two waves as described by \cite{masset97}. In addition to the four-armed beat wave resulting from this coupling, on the $m=4$ spectrum shown in Fig.~\ref{fig:gSa_omega2} we also see a major contribution corresponding to the first harmonic of the bar at $2\omega_{bar}$ (not well shown on this scale, since we are more interested in the outer disc structure), as well as two other well defined features present at all times. All three $m=4$ waves with frequencies lower than that of the bar, run between their 4:1 ILR and CR, slow down at later times, consequently reaching larger radii. The 4:1 ILR of an outer pattern is always closer to the CR of an inner one, i.e., there exists a chain of CR-4:1 ILR overlaps, where an inner wave drives an outer one by non-linear coupling, consistent with the findings of \cite{rautiainen99}. 

In contrast to $m=2$ and 4, the inner $m=3$ mode (and possibly the $m=1$) extend beyond its CR (close to the bar's OLR) (although the amplitude does drop outside the CR, especially at the later times). Nevertheless, the outer/slower three-armed pattern has its 3:1 (first order) ILR very close to the CR of the inner/faster three-armed wave at all times, as seen in Fig.~\ref{fig:gSa_omega2}. In fact, we observe a third $m=3$ pattern in the better defined pattern speeds at the bottom left panel of Fig.~\ref{fig:gSa0_omega} ($r\approx15$~kpc). This occurs at the earlier times when the waves are faster, thus allowing for an additional slower mode. There appears to exist a chain of 3:1 OLR-ILR resonance coupling. That figure also allows to see better that coupling occurs between the inner waves' CR and the outer ones' 3:1 ILR.

It is hard to search for such coupling in the $m=1$ waves due to their low frequency and noisier spectrograms. However, they are seen to follow the evolution of the $m=3$ and $m=4$ and most likely also interacting. 

\section{Migration mechanisms: the effect of corotation versus non-linear mode coupling}
\label{sec:mechanisms}

We would like to know the causes for the strong radial migration seen in the gSa simulation, for which we find the strongest outward transfer of angular momentum.

\subsection{Wide resonances widths in the outer disc}

We showed in the previous Section that the gSa disc presents a system of non-linearly coupled modes. As already discussed, these are expected to result in strong transfer of energy and angular momentum at the coupling regions. To see this effect, in the sixth row of Fig.~\ref{fig:gSa_omega3} we plot the changes in angular momentum for the time periods used to construct the spectrograms in each corresponding column. Due to the large number of modes overlapping, it is not an easy task to relate each peak in the $L-\Delta L$ plane with a particular resonance region. However, since the number of waves decreases in the outer parts of the disc, we can associate clumps in the $L-\Delta L$ plane with the $m=3$ and 1 modes seen in the same location (indicated by the green vertical lines). Because at such distances from the galactic center the resonance curves flatten, they run almost parallel to a density wave reaching that radius. Consequently, we expect that resonance radii in the outer disc are much larger than in the inner disc. We indeed observe a significantly large, broad clump in the changes of angular momentum in the region $r=15-20$~kpc, indicated by the green-shaded patch. The outermost $m=3$ and 1 density waves  seen in the spectrograms have approximately the same radial extent as the feature in the changes in angular momentum: $\sim5$~kpc. Note that this clump becomes broader as it reaches larger radii, just as expected if resonance widths were to increase. Because we see this effect on a timescale smaller than the long dynamical times in the outer disc, we expect that the effective resonance here is CR. Remarkably, since the the outermost $m=3$ spiral is mostly inside its CR, stars are preferentially shifted outward. 

\subsection{The $L-\Delta L$ plane near the corotation resonance}

We found in the previous section that only the $m=3$ wave crosses the CR radius, while it still peaks in amplitude inside it. The lack of structure extending to the OLR may be related to the coupling with the bar since it was also found by \cite{quillen09}. Given this important detail in the type of structure we find in our simulations, it is already clear that migration has to occur in a different way compared to the mechanism of transient spiral structure, where a spiral wave extends between its ILR and OLR. As first shown by \cite{sellwood02}, such a wave results in a particular shape of the changes in angular momentum in the $L-\Delta L$ plane, which is a ridge of negative slope, very similar to what is seen near the bar's CR in our plots (which, incidentally, does not cross its CR). As shown in a number of works \citep{sellwood02,mf10,roskar11}, this feature has 180$^\circ$ rotational symmetry around the point defined by the intersection of the line $\Delta L=0$ and the vertical defining the corotation radius. Therefore, there is an increase in $\Delta L$ interior to the corotation radius and a decrease exterior to it. 

To make this more clear, in Fig.~\ref{fig:mf10} we reproduce part of Fig.~2 by \cite{mf10} 
and shows the $L-\Delta L$ plane for a single perturber: a bar (left panel) and a four-armed spiral wave (right panel). The bar and spiral pattern speeds and strengths are indicated by $\Omega_{b,s}$ and $\epsilon_{b,s}$, respectively, and correspond roughly to the ones seen in the gSa model. This shape of the distributions, with an increase in $\Delta L$ inside and a decrease outside the CR (indicated by the green lines) is characteristic for a single perturber. If all migration came from the CR, i.e., waves were not interacting, then a collection of waves propagating at the same time would result in a series of such symmetric shapes. However, when non-linear exchange of angular momentum occurs, such as due to mode coupling (e.g., \citealt{tagger87, masset97}), structure in $L-\Delta L$ may be quite different, as shown by \cite{mf10}. 

\begin{figure}
\centering
\includegraphics[width=8cm]{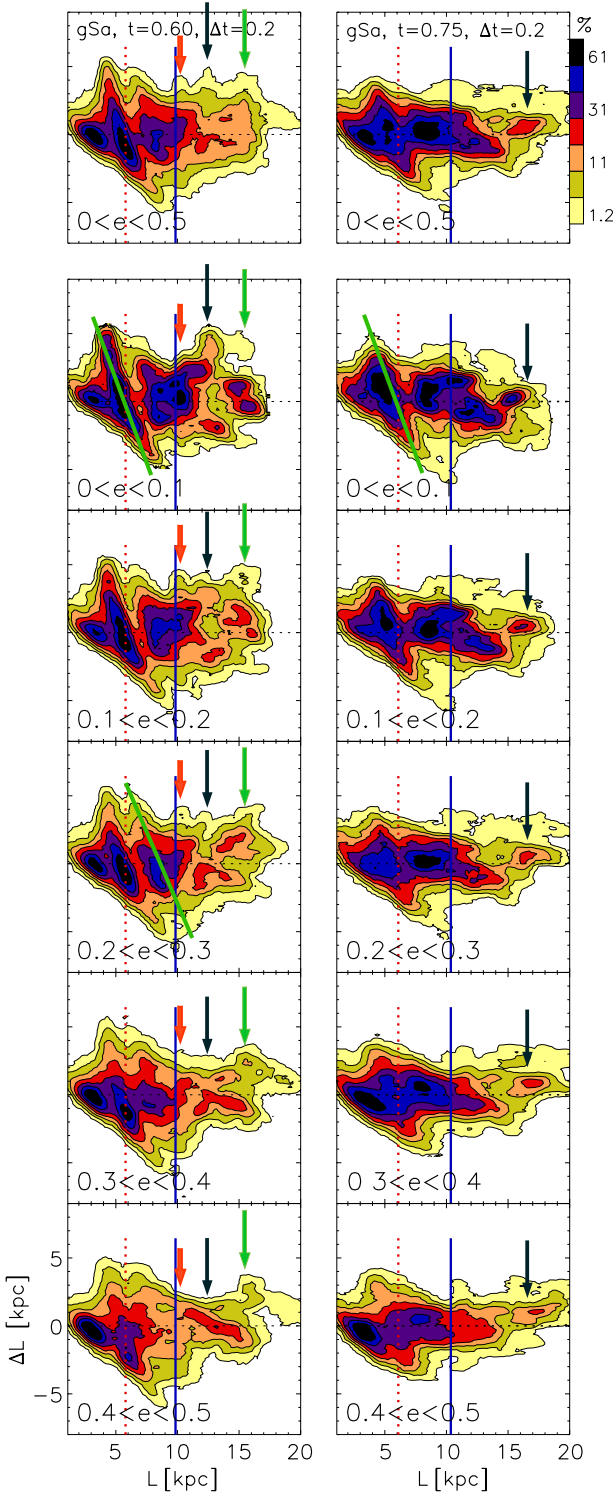}
\caption{
Comparison between the changes in angular momentum for particle samples with different eccentricities at two distinct times for the gSa model. Eccentricities are measured {\it after} the particles have migrated. {\bf First row:} Incremental changes in $\Delta L$ in 200~Myr, centered on 0.6 (left) and 0.75~Gyr (right).  {\bf Rows 2-7:} Subsamples from the total sample shown in the first row, for particles with increasing eccentricities, as shown in each panel. There are no overlaps in the subsets. The green lines of negative slope show three CR-like features. The only well defined such structure seen outside the bar is found in the fourth row. Note that cold orbits in the outer disc (second row) do not exhibit the behavior expected at a CR (clearly seen for the bar), indicating the effects of non-linear coupling among waves. The three prominent peaks shown by the arrows in the upper left panel change differently with variations in eccentricity, suggesting that different dynamical processes are responsible for them.  
\label{fig:gSa_dL_sig}
}
\end{figure}

\subsection{Changes in angular momentum for cold and hot orbits}

\begin{figure*}
\centering
\includegraphics[width=15cm]{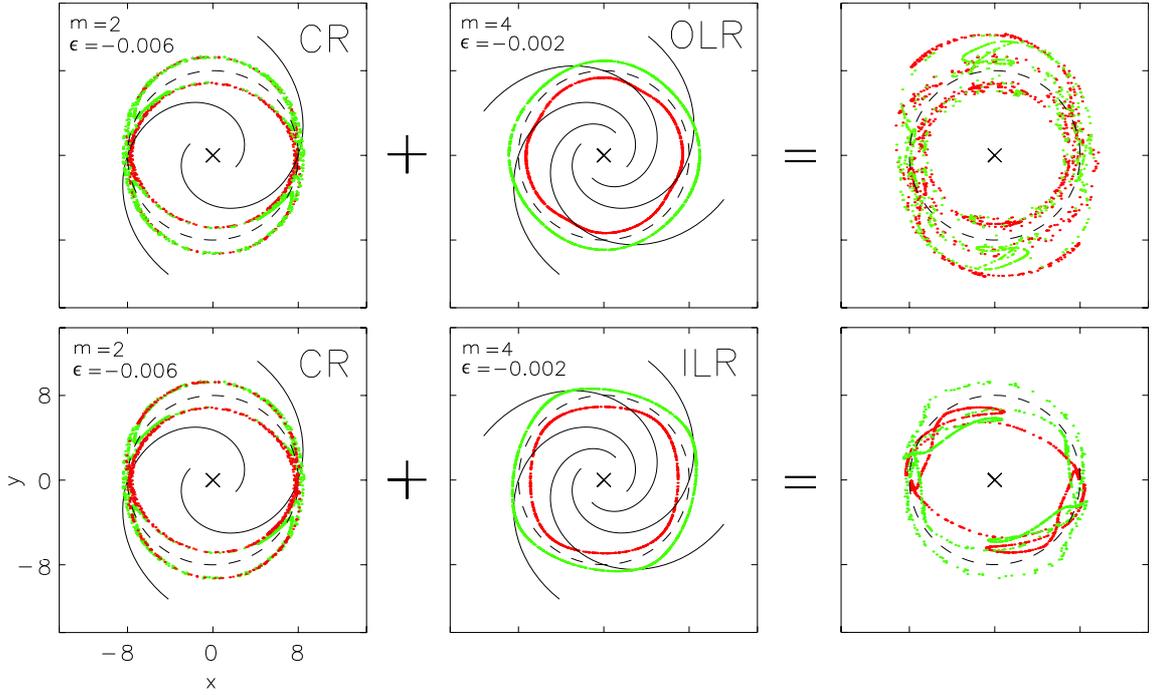}
\caption{ 
Destruction of the horseshoe orbits near the CR in the vicinity of an ILR or OLR of a secondary wave. {\bf First column:} The red/green particles initially start on rings just inside/outside the CR (dashed black circle) of an $m=2$ spiral wave. The top and bottom plots are identical. {\bf Second column:} Same initial conditions, but near the 4:1 OLR (top) or 4:1 ILR (bottom) of an $m=4$ wave. {\bf Third column:} Same initial conditions, but particles are perturbed by both spiral waves. Note that a secondary wave of only 1/3 the strength of the first one is enough to disrupt the horseshoe orbits near the CR. Simulations from \cite{mq06}.
\label{fig:mq06}
}
\end{figure*}

The coupling between waves we found in Section~\ref{sec:coupling} and the complex structure in the $L-\Delta L$ plane makes it compelling to conclude that the migration in our simulations is dominated by resonance overlap associated with multiple patterns, rather than the pure effect of the CR. However, it may be possible that the majority of migration is done by the CR of spiral structure but the additional contribution from non-linear mechanisms contaminates the $L-\Delta L$ space. We can check this, since particles which migrate due to the CR presumably end up on cold orbits \citep{sellwood02,roskar11} and, thus, should be easy to identify. Therefore, in Fig.~\ref{fig:gSa_dL_sig} we compare the structure in the changes in angular momentum for stellar particle samples with different eccentricities at two distinct times for the gSa model. Similarly to Figs.~\ref{fig:gSall_am} and \ref{fig:gSa_omega3}, here in the first row we show the changes in $\Delta L$ in 200~Myr centered on 0.6 (left) and 0.75~Gyr (right). We next estimate the particle eccentricities as 
\begin{equation}\label{eq:e}
e = \sqrt{\frac{u^2+\frac{\kappa^2}
{\Omega^2}v^2}{r^2\kappa^2}}\,,
\end{equation} 
where $u$ and $v\equiv V_c-V_\phi$ are the residual radial and tangential velocities, respectively, while $\kappa$ and $\Omega$ are the radial epicyclic frequency and the angular frequency. Eq.~\ref{eq:e} is the same as Eq.~8 by \cite{arifyanto06}, but in slightly different notation. For each column, we select subsamples from the total sample shown in the first row, for particles with increasing eccentricities, $e$, from zero to 0.5 in units of 0.1, as indicated in each panel. We only consider particles with $e<0.5$, which comprise $\sim85\%$ of the total sample. Therefore, the ``total" distribution shown in each first-row panel is the sum of all panels below. There are no overlaps in the subsets. Note, that we estimate eccentricities {\it after} the particles have migrated, therefore, Fig.~\ref{fig:gSa_dL_sig} does not tell us whether particles originated from a kinematically hot or cold population. 

The panels in the second row of Fig.~\ref{fig:gSa_dL_sig} show particle distributions in the range $0<e<0.1$. As discussed above, for these cold orbits we expect to see the well-defined ridges of negative slope in the $L-\Delta L$ plane, if migration were dominated by the spirals' CR. This expectation is borne out only near the bar's CR (shown by the green lines), more symmetric in the left plot, although some distortion is apparent at negative $\Delta L$. Nevertheless, for the coldest subsample, the bar's CR is the only location resembling what is expected to be the effect of CR. This is somewhat surprising in view of the theory developed by \cite{sellwood02}, where for effective mixing a density wave has to be transient in amplitude so that stars near the CR (on horseshoe orbits) remain trapped away from their initial guiding radii, once the wave has disappeared. While orbits near the bar's CR are similar to those near the CR of a spiral wave, our bars remain stable and only slowly evolve with time in both amplitude and pattern speed (see Fig.~\ref{fig:om}). As we can see in all $L-\Delta L$ plots presented so far, the bar's CR is effective for all of our models in the 3~Gyr of the time evolution. This fact renders the bars in our simulations the most effective drivers of radial migration, despite the fact that they are not transient, but only slowly evolving.

We now examine structure in the $L-\Delta L$ plane for the increasingly hotter subsamples in Fig.~\ref{fig:gSa_dL_sig}. The red, black and green arrows in the left column (lined up in the subsequent rows) show the location of three of the most prominent clumps with positive $\Delta L$. We find a continuously decreasing fraction of particles at the location indicated by the black arrow, with increasing eccentricities, while the opposite effect is seen for the green arrow. Interesting, this $L$-position of the green arrow turns out to be very close to the minimum in the distribution at the hottest samples (fifth row, left). The peak in $\Delta L$, indicated by the red arrow, also increases with eccentricity and peaks at $0.2<e<0.3$. Given this different behavior, the three prominent peaks in the top left panel of Fig.~\ref{fig:gSa_dL_sig} are most likely related to distinct dynamical processes. 

An interesting feature is also found at the later time in the simulation (right panel of Fig.~\ref{fig:gSa_dL_sig}). The (aligned) black arrows show the $L$-position of the densest outer clump seen in the total distribution. This feature is progressively shifted to outer radii by about 3~kpc as the eccentricities grow form 0.1 to 0.5. Additionally, this clump, initially centered slightly above $\Delta L=0$, moves to positive $\Delta L$ with increasing eccentricities. Similar displacement for hot orbits is seen in the first column, as well. What this tells us is that particles migrating outward end up on hot orbits (compared to the local population). Examining the coldest subsamples, we note that the opposite is true for inward migrators. This behavior is related to the approximate conservation of radial action of migrating stars as discussed in detail by \cite{minchev12b}. In Section~\ref{sec:den_sp} we related these outer clumps in the $L-\Delta L$ plane to the CR of $m=1$ and 3 modes. If this is true, then the CR can deliver stars on hot orbits as well, contrary to the expectations.

\begin{figure*}
\centering
\includegraphics[width=16cm]{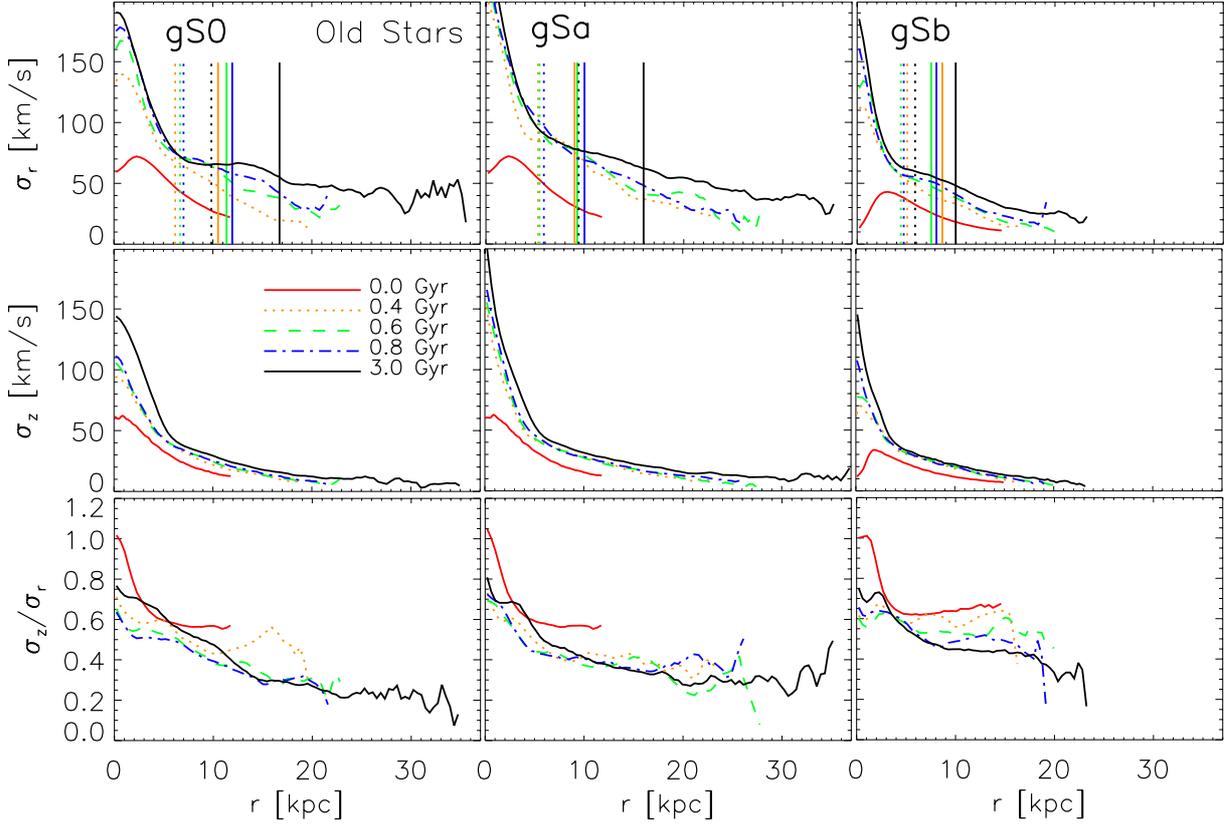}
\caption{
{\bf First row:} Time evolution of the radial velocity dispersion profiles, $\sigma_r(r)$, for the gS0, gSa, and gSb models. Different colors and line styles correspond to the times indicated in the second row, first panel. The bars' CR and 2:1 OLR are shown by the dotted and solid vertical lines, respectively. {\bf Second row:} Same but for the vertical velocity dispersion, $\sigma_z(r)$. {\bf Third row:} Same but for the ratio $\sigma_z/\sigma_r$. The slope in $\sigma_z/\sigma_r$ outside the bar shows that the radial profile flattens more with time than the vertical one, while its overall decrease means that the disc heats more radially than vertically. This is related to the conservation of vertical action. See text for discussion. 
\label{fig:gSall_sig}
}
\end{figure*}

The only CR-like feature away from the bar's CR is found in the left column indicated by the slanted solid green line in the fourth row. It is interesting that this structure is smoother for orbits with $e>0.1$, rather than the cold ones. We now look for the wave responsible for that. Inspecting Fig.~\ref{fig:gSa_omega3}, we find that the nearest CR at this time step (600~Myr) is that of the $m=2$ inner spiral. As discussed earlier, when animated, this wave bounces between the slow, outer $m=2$ spiral and the bar, thus sweeping its CR through a significant portion of the disc. In the particular time depicted here, we happen to catch all its dynamical effect: we see it right between the bar and the outer spiral in the spectrograms and the time window for making a spectrogram is $\pm150$~Myr, i.e., longer than its lifetime. The relation between this density wave and the changes in angular momentum at the same radius is indicated in Fig.~\ref{fig:gSa_omega3}, middle column, by the long green vertical line connecting the alleged cause and effect. The next nearest CR is that of the inner $m=3$ wave, but it is situated outside the bar's 2:1 OLR (blue vertical). There is no CR-like feature in the changes of angular momentum in Fig.~\ref{fig:gSa_omega3} associated with that. We, thus, conclude that the feature outside the bar's CR (shown by a green slanted line in both Figs.~\ref{fig:gSa_omega3} and \ref{fig:gSa_dL_sig}) is caused by the resonant sweeping of the faster $m=2$ wave. In other words, neither in the cold nor hot orbits do we find the signature of transient spirals of constant pattern speed, which indicates that interaction of multiple patterns is important, as we elaborate further in the next Section.

While structure in the changes of angular momentum in the outer disc varies significantly with changing the particles' eccentricities, the effect of the bar's CR seems to be similar in all plots, i.e., the bar's CR is able to migrate stars on both cold and hot orbits with similar efficiency.

\subsection{Destruction of the horseshoe orbits near the corotation resonance}

We showed in Fig.~\ref{fig:mf10} that near the CR both the bar and spiral waves (as single perturbers in test-particle simulations) give rise to a line of negative slope in the $L-\Delta L$ plane. However, we showed in Fig.~\ref{fig:gSa_dL_sig} that even for cold orbits this shape was not seen for any of the spiral waves, except possibly for the one with the variable pattern speed (discussed at the end of the previous Section). What could be the reason for the lack of this expected behavior? It is easy to realize that the shape of structure in the $L-\Delta L$ plane is related to the shape of the particles' trajectories. Therefore, we now look at the behavior of these ``horseshoe" orbits near the CR in the presence of only one and then two spiral waves. 

In a study of the effect of multiple spiral waves on the dynamics of galactic discs, \cite{mq06} showed that in spatial regions where resonances overlap, particle motions become stochastic. This is illustrated in Fig.~\ref{fig:mq06}, where we reproduce a combination of Figs.~9-11 by \cite{mq06}. The strength of the spiral waves is given by the values of $\epsilon$ in the plots, in units as described by \cite{mq06}. Both of these are weaker than the estimated strength of the MW spiral structure (see discussion in \citealt{mf10}).

The first column shows a face-on view of particles starting initially on a ring just inside (red) and a ring just outside (green) the CR (dashed black circle) of an $m=2$ spiral wave of an intermediate strength. Top and bottom plots are identical for the first column. The efficiency of the CR to migrate stars is impressive: stars from the inner (outer) ring are shifted outwards (inwards) by about 3~kpc. Note the unbroken, banana, or ``horseshoe", shape of the orbits. 

The second column presents the same initial conditions, but near the 4:1 OLR (top) or the 4:1 ILR (bottom) of an $m=4$ wave. Here only small distortion is apparent, due to (1) the week spiral amplitude and (2) the fact that near Lindblad resonances stars mostly increase their eccentricities by hardly affect angular momenta. 

Finally, in the third column the two rings of particles are perturbed by both spiral waves shown in each row. Both top and bottom panels are in the reference frame of the corotating, $m=2$ spiral. Remarkably, the secondary wave, which has an amplitude of only 1/3 that of the first one, is enough to disrupt the horseshoe orbits near the CR. Moreover, the distances travelled by the particles are now much larger, especially in the first row. The irregular shapes of these orbits suggests the presence of stochasticity. Such disruption in the horseshoe orbits can naturally explain the lack of structure in the $L-\Delta L$ plane, expected for the pure effect of CR.

We have just shown that a secondary, weaker spiral, is already enough to increase dramatically the effect of CR of the primary, corotating spiral. In our simulations we have not two, but a much larger number of waves overlapping at any given radial range. We conclude, therefore, that the strong migration seen in our simulation comes from non-linear effects, rather than the sole effect of the CR.

\section{Velocity dispersions and disc thickness}
\label{sec:disp}

Previous theoretical studies that focussed on radial migration have not discussed the evolution of disc velocity dispersion profiles nor disc thickness. However, knowledge of these properties is important as the spiral and bar instabilities, which drive the mass build-up in the outer disc, are also expected to be effective at disc heating in both radial and vertical directions. When perturbations at two (or more) pattern speeds, such as a bar and a spiral density wave are present in the disc, the stellar dynamics can be stochastic, particularly near resonances associated with one of the patterns \citep{quillen03,mq06}. 

In Fig.~\ref{fig:gSall_sig} we show the time evolution of the radial, $\sigma_r(r)$, and vertical, $\sigma_z(r)$, velocity dispersion profiles for the old stellar populations of the gS0, gSa, and gSb models in the first, second and third columns, respectively. Initial bulge components are not considered. In each panel the different color curves show different times, as indicated in the bottom left panel. The dotted and solid vertical lines in the first row show the radial positions of the bar's CR and OLR with colors matching the different times.

We first compare the $\sigma_r$ profiles of the gS0 and gSa models, reminding the reader that these two simulations start with the same initial conditions except for the initial 10\% gaseous disc for gSa. We note some interesting differences. Firstly, the gSa's disc is hotter at all radii and all times after bar formation, significantly so in the inner disc. The stronger heating inside the bar's CR of gSa is due to the central mass concentration coming from gas inflow inside the bar following its formation, (e.g., \citealt{friedli93}, see also the gas density profile, third row of Fig.~\ref{fig:gSall}). While this simulation has a weaker bar than the dissipationless gS0, its stronger spirals (see Fig.~\ref{fig:gSall_am}) results in the larger increase of $\sigma_r$ in the outer disc, as well. It is notable that the vertical velocity dispersions increase much less than the radial ones (true for all models). For instance, contrast this difference for gS0 at $r\sim12$~kpc: $\sim$~45 km/s (a factor of three) increase in $\sigma_r$ and $\sim$~12 km/s (a factor of two) increase in $\sigma_z$. 

A remarkable feature in both the gS0 and gSa radial velocity dispersion profiles is the flattening at later times with pivot points at their CR. Similar changes are also seen in the vertical velocity dispersion profiles, however the effect is significantly stronger for $\sigma_r(r)$. This is evident in the ratio $\sigma_z/\sigma_r$ shown in the bottom row of Fig.~\ref{fig:gSall_sig}, where a decline with radius is seen, especially for gS0. We have recently described this phenomenon in \cite{minchev11b,minchev12b}, relating it to the conservation of radial and vertical actions of particles as they migrate. In a simple galactic disc model these can be written as $J_z = E_z/\nu$ and $J_p=E_p/\kappa$, respectively. As stars migrate, they roughly preserve their actions but not their corresponding energies, which vary with radius as the radial and vertical epicyclic frequencies, $\kappa$ and $\nu$, respectively. Thus, as stars migrate inwards (outwards) they heat up (cool down). Stellar radial migration also mixes the radial distributions of the actions $\langle J_p\rangle$ and $\langle J_z\rangle$: the more the disc mixes, the weaker their variation with radius. Therefore, even if the migration rate remains constant, for a coeval stellar population we can expect less heating due to migration at later times. We can see that this is indeed the case by observing that in the region of strongest migration, $r<2.5h_d\sim10$~kpc, velocity dispersion increase saturates quickly with time, while the disc outskirts continuously heat since all the stars populating them are outward migrators (with larger actions).      

The gSb model also exhibits flattening in the $\sigma_r$ and $\sigma_z$ profiles, but not as pronounced due to its weaker bar and spirals and, thus, mixing.

\begin{figure}
\includegraphics[width=9cm]{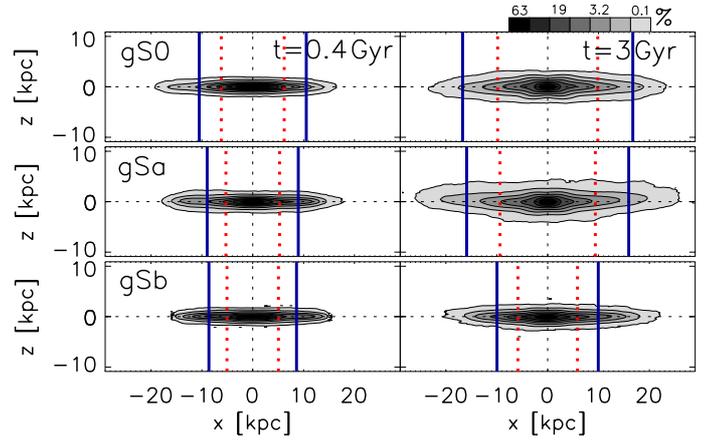}
\caption{
Edge-on stellar density contours for the gS0, gSa, and gSb models at $t=0.4$ and 3~Gyr. The dotted-red and solid-blue vertical lines show the bars' CR and OLR. Note that as the discs extend they also thicken. 
\label{fig:gSall_xz}
}
\end{figure}

We now consider the disc thickening in the three models discussed so far. In Fig.~\ref{fig:gSall_xz} we display edge-on density contour plots for $t=0.4$ and 3~Gyr in the left and right columns, respectively. The CR and OLR are indicated by the dotted red and solid blue vertical lines. Note that as the disc extends it also thickens considerably, as expected from the increase in $\sigma_z$ we showed in Fig.~\ref{fig:gSall_sig}. The increase in scale-height at 2-3 disc scale-lengths is double the initial thickness ($\sim300$~pc) for gS0 and gSa and increases by a factor of 1.5 for gSb. Note that the disc subjected to the strongest migration thickens the most (gSa in this case) as expected from our earlier discussion in this Section. Nevertheless, the disc thickness is not sufficient to be considered a thick disc, with the additional inconsistency of flaring at large radii. This has led \cite{minchev12b} to the conclusion that radial migration cannot result in the formation of thick discs and most recently confirmed by \cite{minchev12c} for a MW-like simulation in the cosmological context.

An important consequence of the results of this Section is that if extended discs originate through migration, then disc outskirts must be kinematically hot.

\section{The effect of gas accretion on the radial disc profiles}
\label{sec:gas_accr}

\begin{figure}
\includegraphics[width=9cm]{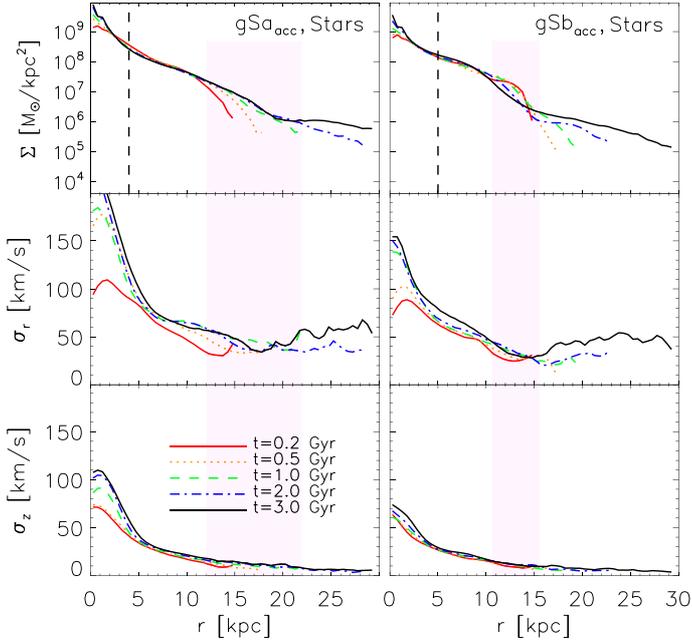}
\caption{
The effect of smooth, in-plane gas accretion. The pink strips show the gas accretion region for simulations starting with the initial conditions used for the gSa and gSb models. The radial disc profiles are shown in the first row. An up-turn in the radial velocity dispersion profile, $\sigma_r(r)$, close to the disc break is evident in the second row, while the vertical velocity dispersion continuously decrease with $r$ (bottom row).
\label{fig:acc}
}
\end{figure}

So far in this paper we have neglected the effect of mergers and gas accretion by filaments. Studying the effect of these would require to explore a large variety of galaxy formation models in a cosmological context, investigating how the results would depend on changing the rate and radial extent of accretion. While we defer such a detailed study to another work, in this Section we briefly explore how our results change when smooth, in-plane gas accretion is considered for the gSa and gSb models, which we refer to as gSa$_{acc}$ and gSb$_{acc}$, respectively. Description of the simulation technique was given in Sec.~\ref{sec:acc} and parameter for both runs can be found in Table~\ref{parameters}. We present our results in Fig.~\ref{fig:acc}. Starting at the initial time, gas is being accreted at the constant rate of 5~$M_\odot$/yr in the range indicated by the pink strip in each column. Varying the extent of this radial range does not change the qualitative results. 

Fig.~\ref{fig:acc} presents the changes in the azimuthally averaged stellar density profiles. Initially the discs extend in Type II profiles, similarly to the gSa and gSb models, shown in Fig.~\ref{fig:gSall}. However, after $\sim 2$~Gyr of gas inflow, the outer stellar discs develop Type III (up-turning) profiles. While the stellar density shown in Fig.~\ref{fig:acc} contains some contribution from the newly formed stars, we have checked that the effect on the old stellar population alone is negligible.  We also note that the total (stars + gas) density profiles (not shown) are very close to Type I. The stars and gas appear to rearrange in such a way as to preserve a single exponential for the total baryonic density.

The middle and bottom rows of Fig.~\ref{fig:acc} show the effect on the radial, $\sigma_r(r)$, and vertical, $\sigma_z(r)$, velocity dispersion radial profiles. It is remarkable that $\sigma_r(r)$ develops an up-turn in the vicinity of the outermost break in the density profiles. In contrast, $\sigma_z$ at all times continues to decrease with radius. This indicates that the stars forming the disc outskirts are on highly eccentric orbits, but well confined to the disc plane. We have to note that this may be a result of the in-plane accreting we consider here.
  
\section{Comparison to observed radial disc profiles}
\label{sec:match}

\subsection{Type I profiles}

Type I profiles are single-exponential profile. In the classification by EPB08, these extend to at least four disc scale-lengths. In terms of bar lengths, the statistics for $r_{\rm break}/r_{\rm bar}(A_{2,max})$ for the S0--Sb galaxies of \cite{erwin08} and \cite{gutierrez11}, are Median = 4.6 and Mean = $5.7\pm 2.5$ (min = 3.3, max = 14.4). In our simulations a single exponential can be fit to the gSa radial profile out to $\sim25$~kpc (Fig.~\ref{fig:gSall}, top row, middle column), which is about five disc scale-lengths at the final time, or $\sim5.5$ bar lengths, considering the gSa $A_{2,max}\sim4.5$ (Fig.~\ref{fig:bars}). Therefore, our gSa model can be considered a case of Type I.

However, we note that while it is possible that the gSa profile could be mistaken for a Type I in the limit of a low S/N image that prevents measuring the profile very far out, there still remain a number of Type I profiles extending to much larger radii, which cannot be explained with this model.

It has been pointed out by \cite{schaye04} that the main problem for the star formation threshold as an explanation for profile breaks is that it fails to explain purely exponential galaxies out to large radii. While here we show that radial mixing can be the driver for these, it should be kept in mind that stars arriving at the disc outskirts by migrating over large distances will be kinematically hot as shown in Fig.~\ref{fig:gSall_sig}. 

\subsection{Type II profiles}

 All of the galaxy models presented in Fig.~\ref{fig:gSall} are consistent with Type II profiles: a single break beyond which the exponential steepens. There are three observed flavors of this type, as described by EPB08: 
\newline
(i) The break can happen at the bar's OLR (Type II.o--OLR). An example of this is found in our gS0 model in agreement with \cite{debattista06}. 
\newline
(ii) The disc break has also been observed to occur outside the bar's OLR (Type II.o--CT). These profiles are currently only explained by star-formation thresholds (e.g., \citealt{kennicutt89, schaye04, elmegreen06}). We now show that our gSa model results in such a profile naturally, as a result of the strong angular momentum outward transfer. It is impressive that the break is more than two scale-lengths beyond the bar's OLR after three Gyr of evolution. The final state of the gSb disc also exhibits a Type II.o--CT profile, occurring at the initial disc truncation radius.
\newline
(iii) The third kind is the Type II.i profile, for which the break occurs at or interior to the end of the bar. We do not find such an example in our simulations. 

\subsection{Type III profiles}

Type III profiles exhibit a shallower exponential beyond the break radius, i.e., the disc density drops-off slower with radius than in the inner disc (the disc is ``antitruncated"). One possibility is that these result from gas-rich, minor mergers, as shown by \cite{younger07}.

We now show for the first time, that Type III profiles can also form in a smooth gas-accretion event, such as from cosmic filaments (Fig.~\ref{fig:acc}). The gSa$_{acc}$ appears consistent with a pure Type III, although both the inner and outer segments are not fitted very well by single exponentials. On the other hand, gSb$_{acc}$ is a case of a Type II + III, composite profile. However, it is clear that, depending on the gas-accretion radius and properties of the bar and spiral structure, a pure Type III profile can result. We, therefore, suggest that observed Type III profiles, combined with high velocity dispersion in the up-turning region, could be a signature of gas accretion.  

An interesting result from our gas-accretion models is the Type III and Type II breaks for the existing stellar population and the gas, respectively, at the accretion region, while the overall baryonic density remains roughly exponential. It is possible that this can later evolve into a Type I or a Type II stellar profile.

The correlation between Type IIIs and unbarred or weakly barred galaxies (EPB08) is consistent with the model presented here, as it is expected that bars are weakened (e.g., citealt{athanassoula05,debattista06}) and even destroyed \citep{bournaud02,combes11} in a gas accretion event.

\subsection{Composite profiles}

Composite profiles are profiles composed of several different types. These are not uncommon. For instance, EPB08 and \cite{gutierrez11} found 6\% (barred) and 8\% (both barred and unbarred) of there samples, respectively, to exhibit Type II+III profiles, e.g., NGC~3982 in EPB08, or NGC~3455 and 5273 in \cite{gutierrez11}. An example of a Type II+III in our simulations can be found in the gSb$_{acc}$ model (Fig.~\ref{fig:acc}). We again stress the fact the stars beyond the outermost disc break are expected to be kinematically hot (see Fig.~\ref{fig:acc}, third panel). 

\section{Discussion and conclusions} 

We have investigated the time evolution of initially truncated galactic discs, via Tree-SPH N-body simulations. We find that due to radial migration and torques associated with bar and spiral instabilities, discs can triple their initial extent, pointing out to a strong annular moment outward transfer. If the bar and spiral structure are sufficiently strong, a single exponential can extend up to $\sim6$ initial scale-lengths ($\sim4$~bar lengths), i.e., doubling the initial disc radius, before it steepens. If resulting from radial migration (as opposed to an in-situ formation), old stellar populations in disc outskirts must exhibit high velocity dispersions, especially in the radial direction, and thus a thickened disc component. Observations of edge-on galaxies, or measurements of the vertical velocities of face-on systems, can be used to distinguish between these two different possibilities. 

In the presence of a central bar, the transfer of stellar mass outward in the disc in the first two disc scale-lengths is dominated by the diffusion resulting from the bar's CR and OLR interacting with resonances associated with the inner spirals (always present in a barred disc). We have shown that the bar's corotation is the most effective driver of radial migration for the extent of our simulations (3~Gyr). For additional migration outwards, a chain of spiral waves, reaching beyond $\sim3h_d$ is needed for angular momentum transport. Self-gravitating spirals at such large radii can result by increasing the stellar density in that region, provided the bar is strong enough (or by gas accretion). A strong bar alone, however, is not sufficient. The stellar population in the outer region needs to be cool enough so that the Toomre instability parameter, $Q_t\sim\sigma_r/\Sigma$, is sufficiently low. Indeed, even though the gS0 model develops stronger bar, the lack of gas results in large velocity dispersion in the region where the outer spirals are needed to carry the angular momentum further out. In contrast, the 10\% gas fraction present initially in the gSa model is enough to prevent this from happening, and thus, spirals extend to radii greater than $4h_d$. Chains of non-linearly coupled spiral waves can develop and carry angular momentum much further out than a single one could (see Section~\ref{sec:coupling}). Note that the gaseous discs are not initially in equilibrium, but this is reached after 100-200~Myr. Even if the disc is initially not completely stable, it is not this initial instability that determines the evolution and migration we discuss in the paper. An indication of this is contained in \cite{minchev11a}, where we examined how the disc dynamics change with increasing the resolution. We showed that although asymmetries developed later (well after the initial relaxation) when a higher number of particles was employed, the overall results were in agreement. This means that independently of the initial relaxation of the system, it is the appearance and presence of strong stellar asymmetries that drives migration (also in agreement with \citealt{brunetti11}).

In the simulations without gas accretion, the single initial exponential disc profile quickly turns into three segments following the formation of a central bar. These can be described roughly by three exponentials (Note that all plots in this paper exclude the initial bulge component).
\newline
(1) The first exponential is formed by steepening the density within the disc scale-length, $h_d$. Increasing the density in the first scale-length is equivalent to building a bulge-like component and is related to the buckling of the bar. The velocity dispersion in those regions are also consistent with bulge kinematics (Fig.~\ref{fig:gSall_sig}). This mass transfer toward the central region is caused by bar torques, angular momentum exchange across the bar's CR and interaction of spiral and bar resonances. 
\newline
(2) The second segment forms by flattening the initial density profile and ends either at the initial truncation radius (gSb), the bar's OLR (gS0), or well outside the bar's OLR (gSa). This exponential continuously flattens with time, i.e., the scale-length increases, as this is where the strongest disc asymmetries occur resulting in the strongest effects of radial migration. We related the outer break of this exponential to discontinuities in the spiral structure and to significant changes in angular momentum due to multiple patterns and torques (Sections~\ref{sec:multi} and \ref{sec:coupling}).
\newline
(3) The third exponential starts either at the initial truncation radius (gSb) or further out (gS0 and gSa), as described above. It can extend to $\sim~10h_d$ if migration is strong enough. We find that the slope of this disc segment is a strong function of time, as is the location of the break at which it starts. The longer the disc is subjected to spiral and bar perturbations, the less steep it becomes (all our models) and/or the farther out the break is displaced (gSa model). A notable feature in outer discs resulting from radial migration is that stellar populations are hot, particularly in the radial direction.

We note that a fourth exponential can result when gas accretion is considered. This outermost segment is shallower than the inner two, as in a Type III profile (see Fig.~\ref{fig:acc}). 

To find out what gives rise to the strong outward transfer of angular momentum we examined in detail the disc dynamics. We first showed that outer breaks and discontinuities in the spiral structure seen in the density plots coincide with significant changes in angular momentum occurring at the same radii (Section~\ref{sec:multi}), suggesting the interplay of resonances associated with multiple patterns. By creating power spectrograms, we showed that multiple waves of multiplicities $m=1,2,3$ and 4 are indeed present in our discs
(Section~\ref{sec:coupling}), with strong evidence for non-linear coupling, as seen in a number of previous works \citep{tagger87,sygnet88,rautiainen99,masset97,quillen11}. Thus, the various waves involved conspire to carry the energy and angular momentum, extracted by the first mode from the inner parts of the disc, much farther out than it alone could. In contrast to strong turbulence, where a large number of modes interact, mode coupling occurs in situations where only a small number of waves or modes can exist, so that each interacts non-linearly with only a few others (e.g., \citealt{davidson72}). This small number of active modes translates into long correlation times, i.e., the quasi-stationary structure found in the simulations ($> 600$~Myr).

At radii larger than $\sim15$~kpc the curves defining the resonances in the power spectrograms (Figs.~\ref{fig:gSa0_omega}-\ref{fig:gSa_omega3}) become increasingly flattened. Therefore, the resonance widths there must be larger than in the inner disc. This may have profound consequences on the formation of extended disc profiles, as stars will be affected over a large range of radii. For these outer regions, the outer spiral's CR of our gSa model spans a wide range of radii, giving rise to effective migration. Remarkably, since this spiral wave is mostly inside its CR, stars are preferentially shifted outwards, as is expected to happen inside the CR. We note that the breaks in the density profiles in our simulations appear near the CR of spirals in agreement with \cite{roskar11}, rather than the spirals' OLR, as suggested by \cite{debattista06}.

We showed in Section~\ref{sec:mechanisms} that the incremental changes in angular momentum (the $L-\Delta L$ plane) exhibit complex behavior, unlike what is expected from the effect of the corotation resonance (CR) alone. Transient spirals have been suggested to migrate stars without increasing their eccentricities. In an effort to separate their effect from that of resonance overlap, we divided the gSa disc into different stellar subsets according to their eccentricities. We expected to find well defined ridges of negative slope in the $L-\Delta L$ plane for the cold population, were the disc mixing dominated by the CR of spiral waves. However, we found no such features but only the complex, clumpy structure related to the non-linear coupling among the bar and spiral waves of different multiplicities. We explained this with the disruption of the horseshoe orbits \citep{mq06} expected near the CR, from resonances associated with other density waves. We conclude that although the CR is very effective at mixing, in the presence of more that one density wave in the disc, one cannot expect a linear superposition of the corresponding CRs, but rather, a network of waves working together.

The effect of the bars' CR in our models is evident throughout the entire simulations time of 3~Gyr (Figs.~\ref{fig:gSall_am} and \ref{fig:3gyr}). This fact renders the bars in our simulation the most effective drivers of radial migration, despite the fact that they are not transient, but only slowly evolving. This is surprising in view of the theory developed by \cite{sellwood02}, where the CR is effective only if structure is transient, but agrees with the work by \cite{brunetti11} and \cite{minchev12c}.

We also considered the effect of gas accretion (Section~\ref{sec:acc}) and found that this results in the formation of Type III profiles in the stellar population. The stars and gas appear to rearrange in such a way as to preserve a single exponential for the total baryonic density. 

\cite{foyle08} found that once a two-component profile is established, (i) the break remains at the same physical location over time and (ii) the outer profile preserves the initial distribution slope. The time for break development depends on the ratio $m_d/\lambda$, where $m_d$ is the ratio of the disc mass fraction and $\lambda$ the halo spin parameter. In contrast to (i) above, in the present work we find strong temporal dependence for our gSa and gS0 models, where the break advances considerably with time over a period of 3~Gyr. We can reconcile this difference by noting that our discs evolve only to 3~Gyr, while \cite{foyle08} study their galaxies to 10~Gyr. We emphasize that the spiral structure is very important for the outward transfer of angular momentum. In the absence of gas accretion, spiral instabilities weaken considerably in the outer discs due to the increase in velocity dispersion, thus preventing further evolution in the disc profiles. For example, while Foyle et al. did not find a break in their model 98 (their Fig.~17) although a bar was formed at $t\approx6$~Gyr, the reason for this is most likely the lack of sufficiently strong spiral structure (at that later evolution time) needed to transport stars to the outer disc. Similarly to our findings, for their model 41 (Fig.~16) the break can be seen to evolve from $r\approx2.5h_d$ at $t=1$~Gyr to $r\approx4.5h_d$ at $t=4$~Gyr. Another difference between \cite{foyle08} and our work is that we find that the outer disc profiles become flatter with time (ii above). This discrepancy is most likely related to the \cite{foyle08}'s initial density distributions extending to larger radii, in contrast to our initially truncated discs. Therefore, changes in their preexisting outer disc profiles may be hard to see.

Our findings are consistent with \cite{perez04} and \cite{trujillo05}, who have shown that observed face-on galaxies at high redshift ($z\approx1$) have breaks at smaller galactocentric radii relative to nearby galaxies. Of course, this can also be related to the disc inside-out growth.

\cite{vandenbosch01} assumed that the distribution of observed (inner) disc scale-lengths would be coupled to the specific angular momentum of the halo. It was pointed out by \cite{debattista06} that the inner disc scale-length evolves considerably over time, as a result of star formation, and cannot be assumed to be representative of the initial profile disc scale-length. While agreeing with this, \cite{foyle08} have argued that the outer disc scale-length remains stable over the discÕs evolution and is a close match to the initial disc scale-length. In view of our findings, where strong changes with time are observed in both the inner and outer profiles, we conclude that both the inner and outer scale-lengths cannot be used as good tracer of the specific angular momentum of the halo. Note that this conclusion is based on our simulations performed in isolation. It remains to be seen whether this is still a true statement in the cosmological context, where there would be a continuous supply of new gas accreting all over the disc. 

Stars migrate along spiral arms and are deposited in the disc outskirts on a timescale consistent with the pattern speed of the outer spiral density wave. This process will progressively deposit increasingly more metal-rich stars from the inner disc in the outskirts. Given the long dynamical times at large distances from the galactic center, we expect that strong azimuthal variations in metallicity might exist in the outer parts of galactic discs. This is the subject of a follow-up work (Minchev et al., In preparation).

While we investigated the effects of secular evolution in barred discs and the effects of gas accretion, our models consist of pre-assembled discs. For example, a mass increase due an accreting thin disc would shrink the older, thicker component and decrease the velocity dispersion we find in the disc outskirts. Therefore, our results may not accurately represent observed galaxies. Rather, our study will be useful in separating the effects of pure dynamical evolution from those of cosmological environment and merging histories. The tidal effects of small satellites can also mix the outer parts of discs of the host galaxy \citep{quillen09,bird12} and possibly extend disc profiles \citep{younger07}. It would be interesting to study the different signatures of this mechanism compared to the effects of secular evolution we present here. 

\acknowledgements

We would like to thank C. Chiappini,  R. de Jong, and K. Freeman for helpful discussions regarding this work. We also thank the anonymous referee for valuable suggestions which have greatly improved the manuscript.

\end{document}